\documentclass[a4paper,11pt]{article}

\usepackage{jheppub}

\usepackage{latexsym, amssymb} 
\usepackage{amsmath}
\usepackage{graphicx}
\newcommand{\lapprox}{%
\mathrel{%
\setbox0=\hbox{$<$}
\raise0.6ex\copy0\kern-\wd0
\lower0.65ex\hbox{$\sim$}
}}
\newcommand{\gapprox}{%
\mathrel{%
\setbox0=\hbox{$>$}
\raise0.6ex\copy0\kern-\wd0
\lower0.65ex\hbox{$\sim$}
}}

\newcommand{\ba}{\begin{array}}
\newcommand{\ea}{\end{array}}
\newcommand{\bd}{\begin{displaymath}}
\newcommand{\ed}{\end{displaymath}}
\newcommand{\be}{\begin{equation}}
\newcommand{\ee}{\end{equation}}
\newcommand{\bea}{\begin{eqnarray}}
\newcommand{\eea}{\end{eqnarray}}

\def\q2 {q^2}

\def\bt{\begin{table}}
\def\et{\end{table}}

\catcode`@=11 
\def \gsim{\mathrel{\mathpalette\@versim>}}
\def \lsim{\mathrel{\mathpalette\@versim<}}
\def \@versim#1#2{\lower0.4ex\vbox{\baselineskip\z@skip\lineskip\z@skip
     \lineskiplimit\z@\ialign{$\m@th#1\hfil##\hfil$%
     \crcr#2\crcr\sim\crcr}}}
\catcode`@=12 

\begin{document}

\title{An updated analysis of radion-higgs mixing in the light of LHC data}

\author[a]{Nishita Desai}
\author[b]{Ushoshi Maitra}
\author[b]{Biswarup Mukhopadhyaya}
\affiliation[a]{{\em Department of Physics and Astronomy},\\ University College London, UK}
\affiliation[b]{{\em Regional Centre for Accelerator-based Particle Physics}\\
Harish-Chandra Research Institute,\\ Chhatnag Road, Jhusi, Allahabad - 211 019, India}
\emailAdd{n.desai@ucl.ac.uk}
\emailAdd{ushoshi@hri.res.in}
\emailAdd{biswarup@hri.res.in}

\preprint{RECAPP-HRI-2013-012, LCTS/2013-14}

\abstract{ 

  We explore the constraints on the parameter space of a
  Randall-Sundrum warped geometry scenario, where a radion field
  arises out of the attempt to stabilise the radius of the extra
  compact spacelike dimension, using the most recent data from higgs
  searches at the Large Hadron Collider (LHC) and the Tevatron.  We
  calculate contributions from both the scalar mass eigenstates
  arising from radion-higgs kinetic mixing in all important search
  channels.  The most important channel to be affected is the decay
  via $WW^{(*)}$, where no invariant mass peak can discern the two
  distinct physical states.  Improving upon the previous studies, we
  perform a full analysis in the $WW^{(*)}$ channel, taking into
  account the effect of various cuts and interference when the two
  scalar are closely spaced.  We examine both cases where the
  experimentally discovered scalar is either 'higgs-like' or
  'radion-like'. The implications of a relatively massive scalar
  decaying into a pair of 125 GeV scalars is also included. Based on a
  global analysis of the current data, including not only a single 125
  GeV scalar but also another one with mass over the range 110 to 600
  GeV, we obtain the up-to-date exclusion contours in the parameter
  space.  Side by side, regions agreeing with the data within 68\% and
  95\% confidence level based on a $\chi^2$-minimisation procedure,
  are also presented.  }

\maketitle

\section{Introduction}

The announced discovery of a boson in the mass range 125-126 GeV, by
both the ATLAS~\cite{Aad:2012tfa} and CMS~\cite{Chatrchyan:2012ufa}
collaborations at the Large Hadron Collider (LHC) experiment, has
naturally generated a lot of enthusiasm among particle physicists. As
of now, the properties of the particle whose signature has been
avowedly noticed are consistent with those of the Standard Model (SM)
higgs boson. However, the present data also leave some scope for it
being a scalar with a certain degree of non-standard behaviour.  The
analysis of such possibilities, both model-independently and on the
basis of specific theoretical scenarios, has consumed rather
substantial efforts in the recent months.

One scenario of particular interest in this context is one with a
warped extra spacelike dimension.  First proposed by Randall and
Sundrum (RS), it has a non-factorizable geometry with an exponential
warp factor \cite{Randall:1999ee}. Furthermore, the extra dimension is
endowed with an $S_1 /Z_2$ orbifold symmetry, with two 3-branes
residing at the orbifold fixed points, the SM fields being confined to
one of the branes (called the `visible brane', at $y = r_c \pi$, where
$r_c$ is the radius of the compact dimension and $y$ is the
co-ordinate along that dimension).  When the warp-factor in the
exponent has a value of about 35, mass parameters of the order of the
Planck scale in the `bulk' get scaled down to the TeV scale on the
visible branch, thus providing a spectacular explanation of the
hierarchy between these two scales.

A bonus in the low-energy phenomenology of this model is the
occurrence of TeV-scale Kaluza-Klein (KK) excitations of the spin-2
graviton on the visible brane, with coupling to the SM fields
suppressed by the TeV scale \cite{Davoudiasl:1999jd,
  Davoudiasl:2000wi, Chang:1999yn}. The mass limit on the lowest
excitation of the graviton in this scenario has already gone beyond 2
TeV (with certain assumptions on the model parameters)
 \cite{ATLAS-CONF-2013-017, Chatrchyan:2013qha}.  However, another
interesting and testable feature of this theory results from the
mechanism introduced to stabilize the radius of the compact dimension,
where the radius is envisioned as arising from the vacuum expectation
value (vev) of a modular field.  This field can be naturally given a
vev by hypothesizing a brane-dependent potential for it, resulting in
a physical field of geometrical origin, popularly called the radion
field, with mass and vev around the electroweak scale, which couples
to the trace of the energy-momentum tensor \cite{Goldberger:1999uk,
  Goldberger:1999un}.  Consistency with general covariance demands the
addition of terms giving rise to mixing between the SM higgs and the
radion \cite{Giudice:2000av, Csaki:2000zn, Csaki:1999mp,
  Dominici:2002jv, Csaki:2007ns}.  Consequently, speculations have
been made on whether the 125-126 GeV state, instead of being a pure SM
higgs, could instead be the radion, or a mixture of the two.

A number of studies have already taken place in this direction, based
on both the `pure radion' and `radion-higgs mixing'
hypotheses~\cite{Mahanta:2000zp, Gunion:2003px, Toharia:2004pv,
  Toharia:2008tm, Frank:2012nb, Rizzo:2002bt, Cheung:2005pg,
  Han:2001xs, Chaichian:2001rq, Battaglia:2003gb,
  Bhattacherjee:2011yh, Goncalves:2010dw, Barger:2011hu,
  Cheung:2011nv, Davoudiasl:2012xd, Low:2012rj, Soa:2012ra,
  Chacko:2012vm, Barger:2011qn, Grzadkowski:2012ng, deSandes:2011zs,
  Kubota:2012in, Ohno:2013sc, Cho:2013mva}.  In the present work, we
perform a global analysis of the available data, assuming that both of
the physical states arising from radion-higgs mixing contribute to the
event rates in various channels.  Using both the 2011 and 2012 data,
we obtain the best fit points in terms of the parameters of the
model. Furthermore, we obtain the 95\% confidence level contours in
the parameter space, which indicate the extent to which new physics
can be accommodated in the light of the available results.  Side by
side, we identify the regions which are disallowed by data in one or
more channels, as obtained from the published 95\% C.L. exclusion
limits on the signal strength, defined as $\mu = \sigma/\sigma_{SM}$,
where $\sigma$ is the predicted cross-section in the relevant channel
for a specific combination of the model parameters, and $\sigma_{SM}$
is the corresponding prediction for the SM higgs boson. The region
that is left after such exclusion can be treated as one where the
presence of a radion-like (higgs-like) scalar is compatible with the
data as of now. A comparison of this region with the 95\%
C.L. contours around the best fit values of the parameters indicates
the viability (or otherwise) of this particular new physics scenario.

Our work improves upon other recent studies based on LHC data
\cite{Kubota:2012in, Grzadkowski:2012ng, deSandes:2011zs, Cho:2013mva}
in a number of ways.  This is the first global analysis, following a
$\chi^2$-minimisation procedure, of radion-higgs mixing, using the
latest available data from 7 and 8 TeV LHC runs to obtain best fit
parameters and significance contours.  We include the possibility of
an additional scalar mass eigenstate coexisting with the 125 GeV
state, with both of them contributing to the final states looked for,
subject to event selection criteria pertaining to the 125 GeV higgs.
While it is unlikely that the contribution from the additional scalar
will be confused with the signal of a 125 GeV scalar in the
$\gamma\gamma$ and $ZZ^{(*)}$ final states (as the reconstructed
invariant mass will point to two distinct resonances), it cannot {\it
  a priori} be ruled out for the $WW^{(*)}$ channel.  The presence of
two neutrinos in the di-lepton final state makes it impossible to
reconstruct the mass of the parent particle and one would therefore
expect some enhancement to the signal strength due to the extra
contribution from the second state which must be estimated by
simulating the effect of the selection cuts used by the correponding
experimental analyses.  This makes the best-fit regions different from
what one finds with the assumptions that the entire contribution in
every channel comes from one scalar resonance only.

Secondly, we also use the strategy of simulating the full cut-based
analysis in restricting the allowed regions from the available upper
limit on $\sigma/\sigma_{SM}$ for an addition scalar with different
mass, demanding not only (a) the extra contribution at 125 GeV be
smaller than the current upper limit, but also (b) the combined
contribution using cuts correponding to the SM higgs search at the
mass of the extra resonance be smaller than the upper limit at that
mass.  Again, this makes a difference mainly in the $WW^{(*)}$
channel. The contribution here (as also in the case of global fits) is
the sum of those from two distinct mass eigenstates, so that the
acceptance of the cuts does not factor out when taking the ratio to
expected SM cross section.

Thirdly, we have taken into account the interference between processes
mediated by radion-higgs mixed mass eigenstates whenever they are
close to each other.  And finally, we have explicitly included
processes where a relatively heavy, radion(higgs)-dominated state
decays into two higgs(radion)-dominated scalars at 125 GeV, each of
which can go to the decay channels searched for.  In a way, this leads
to an additional production mechanism of the 125 GeV state, which we
have felt should be included in a full analysis.

The presentation of our paper is as follows.  We outline the RS model
with higgs-radion mixing in the next section.  The strategy of our
analysis is described in section 3, while section 4 contains the
numerical results. We summarise and conclude in section 5.

\section{The model and its parameters}

\subsection{The minimal Randall-Sundrum model and the radion}

In the minimal version of Randall-Sundrum (RS) model, one has an extra
warped spacelike compact dimension $y = r_c \phi$, where $r_c$ is the
radius of compactification. An $S_1/Z_2$ orbifolding is applied with a
pair of 3-branes at the orbifold fixed points (at $\phi = 0$ and $\phi
= \pi$). Gravity, propagating in the bulk, peaks at the first of these
branes, usually called the Planck (hidden) brane (at $\phi = 0$),
while the SM fields are confined to the visible brane (at $\phi =
\pi$).\footnote{While various modifications, including for example,
  gauge fileds in the bulk have been considered
  \cite{Goldberger:1999wh, Davoudiasl:1999tf, Pomarol:1999ad,
    Agashe:2006hk, Agashe:2004ay, Agashe:2004cp, Agashe:2003zs}, we
  have, however, confined ourselves to the minimal RS scenario.}

The action for the above configuration is given by
\cite{Randall:1999ee}
\begin {eqnarray}
 S & = & S_{gravity} + S_{v} + S_{h} \nonumber \\ 
 S_{gravity} & = & \int d^4x \int _{-\pi} ^{\pi} d\phi \sqrt{-G} \{-\Lambda + 2M_5^3R\} \nonumber \\ 
 S_{v} & = & \int d^4 x \sqrt{-g_{v}} \{{\cal{L}}_{v} - V_{v}\} \nonumber \\ 
 S_{h} & = & \int d^4 x \sqrt{-g_{h}} \{{\cal{L}}_{h} - V_{h}\}  
\end {eqnarray}
 \noindent 
where the subscripts $v$ and $h$ refer to the visible and hidden
branes respectively, $G$ is the determinant of the five dimensional
metric $G_{MN}$ and the metrics on the visible and hidden branes are
given by
\begin{equation}
 g^{v}_{\mu\nu}(x^\mu)  \equiv  G_{\mu\nu}(x^\mu,\phi = \pi) ,   g^{h}_{\mu\nu}(x^\mu)  \equiv  G_{\mu\nu}(x^\mu,\phi = 0)
\end{equation}
\noindent 
the greek indices being representation of (1+3) dimensional
coordinates on the visible (hidden) brane.  $M_{5}$ is the
5-dimensional Planck mass and $\Lambda$ is the bulk cosmological
constant.  $V_{v}$ and $V_{h}$ are the brane tensions of visible and
hidden branes respectively.

The bulk metric obtained after solving Einstein's equations is then
\begin{equation}
ds^2 = e^{-2k|y|} \eta_{\mu\nu} dx^\mu dx^\nu - dy^2
\end{equation}
\noindent 
where $k = \sqrt{\frac{-\Lambda}{24 M_{5}^3}}$
and
\begin{equation}
V_{h} = -V_{v} = 24M_5^3k.
\end{equation}

$M_5$ is related to the 4-dimensional Planck mass $M_{Pl}$ by
\begin{equation}
 M^2_{Pl} = \frac{M^3_{5}}{k}[1 - e^{-2kr_c\pi}]
\end{equation}
 
The 5-dimensional metric consists solely of mass parameters whose
values are around the Planck scale.  For the choice $kr_c \simeq 12$,
which requires barely an order of disparity between the scales $k$ and
$1/r_c$, the mass parameters on the visible brane are suppressed with
respect to the Planck scale by the exponential factor $e^{kr_c\pi}
\simeq 10^{16}$, thus offering a rather appealing explanation of the
hierarchy between the Planck and TeV scales.  The Kaluza-Klein (KK)
decomposition of the graviton on the visible brane leads to a discrete
tower of states, with one massless graviton and a series of TeV-scale
spin-2 particles. The massless graviton couples to all matter fields
with strength $\sim 1/{M_P}$, while the corresponding couplings for
the massive modes (in the TeV range) receive an exponential
enhancement, thus opening up the possibility of observing signals of
the massive gravitons in TeV-scale experiments
~\cite{Davoudiasl:1999jd, Davoudiasl:2000wi, Chang:1999yn}.  Current
experimental limits from the LHC rule out any mass for the lowest
graviton excitation below $1.15(2.47)$ TeV for $k/M_P \le 0.01(0.1)$
\cite{ATLAS-CONF-2013-017}.
 
The radius of compactification $r_c$, was an input by hand in the
original model, however, it can be given a dynamic origin by linking
it to the vev of a $\phi$-independent modulus field, $T(x)$, so that
$r_c = \langle T \rangle$. We can define a new field
\begin{equation}
\varphi(x) = \Lambda_\varphi e^{-k(T(x) - r_c)\pi}
\end{equation} 
with its vev given by $\Lambda_\varphi =
\sqrt{{\frac{24M_{5}^3}{k}}}e^{-k\pi r_c}$.
\noindent

A vev for the modulus field can be dynamically generated if it has a
potential. To generate the potential for $\varphi(x)$, a scalar field
with bulk action is included along with interaction terms on the
hidden and visible branes. The terms on the branes cause the scalar
field to develop a $\phi$-dependent vev. Inserting this solution into
the bulk scalar action and integrating over $\phi$ yields an effective
potential for $\varphi(x)$ of the form
\begin{equation}
  V_{\varphi} (r_c) = k\epsilon v_h ^2 + 4ke^{-4kr_c\pi}(v_v - v_he^{-\epsilon kr_c\pi})^2(1+\epsilon/4) - k\epsilon v_he^{-(4+\epsilon)kr_c\pi}(2v_v - v_he^{-\epsilon kr_c\pi}) 
\end{equation}
where $\epsilon \simeq m^2/4k^2$
\begin{equation}
V(\varphi)  = \frac{k^3}{144M_{5}^6} \varphi^4(v_v - v_h(\frac{\varphi}{\Lambda_\varphi exp(k\pi r_c)})^\epsilon),
\end{equation} 

\noindent
where $v_v$ and $v_h$ are interaction terms on the visible and hidden
branes respectively and by assumption $\epsilon \ll 1$ This new
massive filed $\varphi$ is the radion field, where mass is obtained
from $\frac{\partial^{2}V(\varphi)}{\partial\varphi^{2}}$.
Furthermore, one obtains the minimum of V($\varphi$) for $kr_c \approx
12$ for $ln(\frac{v_v}{v_h}) \sim 1$.

The radion mass, $m_\varphi$, and the vev $\Lambda_\varphi$,
constitute the set of free parameters of the theory in the radion
sector, which now has the distinction of `naturally' generating a
TeV-scale vev on the visible brane.  They have implications on
particle phenomenology within the reach of the LHC.  In particular,
the radion mass may turn out to be a little below a TeV, thus making
the detection of radion somewhat easier that that of the KK mode of
the graviton \cite{Goldberger:1999uk, Goldberger:1999un}.

Integrating over the orbifold coordinates it can be shown that the
radion field couples to the trace of energy-momentum tensor
$(T^{\mu}_{\nu})$. The canonically normalized effective action is
\begin{equation}
 S_\varphi = \int d^4 x \sqrt{-g}[\frac{2M_5^3}{k}(1 - \frac{\varphi^2}{\Lambda_\varphi^2}e^{-2k\pi r_c})R + \frac{1}{2}\partial_\mu\varphi\partial^\mu\varphi - V(\varphi) + (1 - \frac{\varphi}{\Lambda_\varphi})T_\mu^\mu]
\end{equation}
 
It should be noted that, while the radion has couplings that are very
similar to those of the SM higgs, it has additional interaction with
massless gauge boson (photon, gluon) pairs via the trace anomaly
terms.

\subsection{Radion-Higgs mixing}

In addition to the above action, general covariance also allows a
higgs-radion mixing term \cite{Giudice:2000av}, parametrized by the
dimensionless quantity $\xi$.  Such a term couples the higgs field to
the Ricci scalar of the induced metric ($g_{ind}$) on the visible
brane
\begin{equation}
 S = -\xi\int d^4x \sqrt{-g_{ind}}R(g_{ind})H^{\dagger}H
\end{equation}
 where $H = [(v + h)/\sqrt{2},0]$ with $v = 246$ GeV
 
For phenomenological purpose, we are interested in terms in
$T^\mu_\mu$, which are bilinear in the SM fields.  Retaining such
terms only, one has
\begin{equation}
 T^{\mu}_{\mu} = T^{(1)\mu}_{\mu} + T^{(2)\mu}_{\mu}
\end{equation}
with
\begin{eqnarray}
T^{(1)\mu}_{\mu} & = & 6\xi v\Box h \nonumber \\
T^{(2)\mu}_{\mu} & = & (6\xi - 1)\partial_\mu h\partial^\mu h + 6\xi h \Box h + 2 m_h^2 h^2 + m_{ij}\bar{\psi}_i\psi_j - M_v^2V_{A\mu}V^{\mu}_{A}
\end{eqnarray}

$T^{(1)\mu}_\mu$ induces a kinetic mixing between $\varphi$ and h.
After shifting $\varphi$ with respect to its vacuum expectation value
$\Lambda_\varphi$ we obtain
\begin{equation}
{\cal{L}} = -\frac{1}{2} \varphi(\Box + m_\varphi ^2)\varphi - \frac{1}{2} h(\Box + m_h ^2)h - 6 \xi \frac{v}{\Lambda_\varphi}\varphi\Box h
\end{equation}

We confine our study to a region of the
paremeter space where the radion vev $\Lambda_\varphi$ is well above
the vev of the SM higgs.  Besides, it is phenomenologically safe not
to consider $\xi$ with magnitude much above unity, since a large value
may destabilise the geometry itself through
back-reaction. Thus one can make the further
approximation $6\xi \frac{v}{\Lambda_\varphi} << 1$.  In this
approximation, the kinetic energy terms acquire a canonical form under
the basis transformation from $(\varphi, h)$ to $(\varphi^{'},h^{'})$,
such that

\begin{eqnarray}
\varphi & = & (\sin\theta-\sin\rho \cos\theta)h^{'}+(\cos\theta+\sin\rho \sin\theta)\varphi^{'}\nonumber\\ 
 h & = & \cos\rho \cos\theta h^{'}-\cos\rho \sin\theta \varphi^{'}
\end{eqnarray}
\noindent   
where
\begin{equation}
\tan\rho = 6\xi\frac{v}{\Lambda_\varphi}, ~~~~ \tan2\theta = \frac{2\sin\rho m_\varphi^2}{\cos^2\rho(m_{\varphi}^{2} - m_{h}^{2})} 
\label{eqn:theta}
\end{equation}
\noindent
and one ends up with the physical masses 
\begin{equation}
\label{eqn:mass}
m_{\varphi^{'},h^{'}}^2 = \frac{1}{2}\left[(1 + \sin^{2}\rho) m_\varphi^{2} + \cos^{2}\rho m_h^{2} \pm \sqrt{\cos^4\rho (m_\varphi^2 - m_h^2)^2 + 4\sin^2\rho m_\varphi^4}\right] 
\end{equation}

The interactions of $\varphi^{'}$ and $h^{'}$ with fermions ($f$) and
massive gauge bosons ($V$) is given by

\begin{equation}
 {\cal{L}}_{1}  =  \frac{-1}{v}(m_{ij}\bar{\psi_i}\psi_{j} - M_{v}^2 V_{A\mu}V_{A}^{\mu})(A_{h}h{'} + \frac{v}{\Lambda_{\varphi}} A_{\varphi} \varphi^{'})
\end{equation}

As has been mentioned above, the coupling of $\varphi$ to a pair of
gluons also includes the trace anomaly term.  Taking it into account,
the gluon-gluon couplings for both of the mass eigenstates are given
by
\begin{equation}
{\cal{L}}_{2} = \frac{-1}{v}\frac{\alpha_s}{16\pi}G_{\mu\nu}G^{\mu\nu}(B_{h}h^{'} + \frac{v}{\Lambda_{\varphi}} B_{\varphi} \varphi{'})
\end{equation}
while the corresponding Lagrangian for the photon is
\begin{equation}
 {\cal{L}}_{3} = \frac{-1}{v}\frac{\alpha_{EM}}{8\pi}F_{\mu\nu}F^{\mu\nu}(C_{h}h^{'} + \frac{v}{\Lambda_{\varphi}} C_{\varphi} \varphi{'})
\end{equation}
\noindent
where 
\begin{eqnarray}
\nonumber a_{h}^{1} & = & \frac{v}{\Lambda_{\varphi}}(\sin \theta - \sin \rho \cos \theta),\\
\nonumber a_{h}^{2} & = & \cos \rho \cos \theta,\\
\nonumber a_{\varphi}^{1} & = & \cos \theta + \sin \rho \sin \theta,\\
\nonumber a_{\varphi}^{2} & = & \frac{\Lambda_{\varphi}}{v}(\cos \rho \sin \theta),\\
\nonumber A_{h} & = & a_{h}^{1} + a_{h}^{2}, \\
\nonumber A_{\varphi} & = & a_{\varphi}^{1} - a_{\varphi}^{2}, 
\end{eqnarray}
\begin{eqnarray}
\nonumber B_{h} & = & A_{h} F_{1/2}(\tau_{t}) - 2b_{3}a_{h}^{1}, \\
\nonumber B_{\varphi} & = & A_{\varphi} F_{1/2}(\tau_{t}) - 2b_{3}a_{\varphi}^{1}, \\
\nonumber C_{h} & = &  A_{h} (\frac{4}{3}F_{1/2}(\tau_{t}) + F_{1}(\tau_{W})) - (b_{2} + b_{y})a_{h}^{1},\\
\nonumber C_{\varphi} & = &  A_{\varphi} (\frac{4}{3}F_{1/2}(\tau_{t}) + F_{1}(\tau_{W})) - (b_{2} + b_{y})a_{\varphi}^{1} \\
\nonumber \tau_t & = & \frac{4m^{2}_t}{q^2}, \\
\nonumber \tau_W & = & \frac{4m^{2}_W}{q^2}, \\
b_3 & = & 7,~ b_2 = 19/6, ~ b_Y = -41/6.
\end{eqnarray}
\noindent
where $q^2 = m^2_{h^{'}}(m^2_{\varphi^{'}})$ depending on
$h^{'}(\varphi^{'})\rightarrow gg,\gamma\gamma$. $b_{2}, b_{3}$ and
$b_{Y}$ are the SM $\beta$-function coefficients in $SU(3)$ and
$SU(2)\times U(1)_{Y}$ respectively.  $F_{1}(\tau_W)$ and
$F_{1/2}(\tau_t)$ are the form factor for W and top loop
respectively. The form of these functions are
\begin{eqnarray}
\nonumber F_{1/2}(\tau) & = & -2\tau[1 + (1 - \tau)f(\tau)], \\
\nonumber F_{1}(\tau) & = & 2 + 3\tau + 3\tau(2 - \tau)f(\tau), \\
\nonumber f(\tau) & = & [\sin^{-1}(\frac{1}{\sqrt{\tau}})]^2, ~~~~ if~~\tau \geq 1 \\
\nonumber         & = & \frac{1}{4}[ln(\frac{\eta_+}{\eta_{-}}) - \imath\pi]^2,~~~ if~~ \tau<1 \\
\eta_{\pm} & = & 1 \pm \sqrt{ 1 - \tau}.
\end{eqnarray}

The coupling of $\varphi$ to $h$ depends on the Goldberger-Wise
stabilization potential $V(\varphi)$.  On assuming the self-couplings
of $\varphi$ in $V(\varphi)$ to be small, we have
\begin{equation}
 \Gamma(\varphi^{'}\rightarrow h^{'} h^{'}) = \frac{m_{\varphi'}^{3}}{32\pi\Lambda_{\varphi}^2}[1 - 6\xi + 2\frac{m_{h'}^2}{m_{\varphi'}^2}(1 + 6\xi)]^2\sqrt{[1 - 4\frac{m_{h'}^2}{m_{\varphi{'}}^2}]}
\end{equation}

Obviously, all interactions of either physical state are now functions
of $m_{\varphi^{'}}, m_{h^{'}}, \Lambda_{\varphi}$ and $\xi$.  In our
subsequent calculations, we use these as the basic parameters,
obtaining in each case the quantities $m_\varphi, m_h$ by inverting
(Eqn.~\ref{eqn:mass}).  Requiring that the discriminant in
(Eqn.~\ref{eqn:mass}) to remain positive implies a restriction on the
parameter $\xi$ as a function of the remaining three parameters.  This
constitutes a ``theoretically allowed'' region in $\xi$ for given
($m_{h^{'}}$, $m_{\phi^{'}}$, $\Lambda_\varphi$).  Within this region,
we have two solutions corresponding to $m_\varphi > m_h$ and
$m_\varphi < m_h $ in (Eqn.~\ref{eqn:mass}).  In the first case we
have $m_{\varphi^{'}} \rightarrow m_\varphi$ and $m_{h^{'}}
\rightarrow m_{h}$ in the limit $\xi \rightarrow 0$. Exactly the
opposite happens in the other case, with $m_{\varphi^{'}} \rightarrow
m_{h}$ and $m_{h^{'}} \rightarrow m_\varphi $ as $\xi$ approaches
zero.  A further constraint on $\xi$ follows when one requires
$m_\varphi > m_h$.  This is because one has in that case,
\begin{equation}
 m_\varphi^2 - m_h^2 = \frac{\sqrt{D} - \sin^{2}\rho(m_{\varphi^{'}}^2 + m_{h^{'}}^2)}{1 - \sin^{4}\rho}
\end{equation}
where, 
\begin{equation}
 D = (m_{\varphi^{'}}^2 + m_{h^{'}}^2)^2 - 4(1 + \sin^{2}\rho)m_{\varphi^{'}}^2 m_{h^{'}}^2
\end{equation}
One thus ends up with the condition $ \sqrt{D} > \sin^{2}{\rho} (m_{\varphi^{'}}^2 + m_{h^{'}}^2) $, 
thus yielding an additional constraints on $\xi$.

In the other case described above one has 
\begin{equation}
 m_\varphi^2 - m_h^2 = -\frac{\sqrt{D} + \sin^{2}\rho(m_{\varphi^{'}}^2 + m_{h^{'}}^2)}{1 - \sin^{4}\rho}
\end{equation}
which trivially ensures $m_\varphi < m_h$.

We now define the convention for our analysis.  (Eqn.~\ref{eqn:mass})
implies that the lightest state will always be $h'$.  Thus, when
$m_\varphi < m_h$, $h'$ becomes the radion-dominated state
i.e.\ $m_{h'}\rightarrow m_{\varphi}$ when $\xi \rightarrow 0$.  On
the other hand, when $m_\varphi > m_h$, we have \ $m_{h'}\rightarrow
m_{h}$ when $\xi \rightarrow 0$.  Let us label $\varphi^{'}(h^{'})$ as
the mixed radion state $(R)$ if, on setting $\xi= 0$, one recovers
$m_{\varphi^{'}} = m_\varphi~(m_{h^{'}} = m_\varphi)$. The other state
is named the mixed higgs state (H).

Basically, the two interchangeable limits of the states $h^{'}$ and
$\varphi^{'}$ for $\xi = 0$ in the two cases arise from the fact that
the angle $\theta$ in (Eqn.~\ref{eqn:theta}) is 0 or $\pi/2$,
depending on whether $m_\varphi > m_h$ or $m_\varphi < m_h$. Both of
the above mass inequalities are thus implicit in
(Eqn.~\ref{eqn:mass}).

\section{Strategy for analysis}

We propose to scan over the parameter space in terms of masses of the
observable physical eigenstates $m_H$ and $m_R$ for all allowed values
of the mixing parameter $\xi$ for a given $\Lambda_\varphi$. Since one
scalar has been discovered at the LHC, two possibilities arise ---
viz.\ we identify the resonance near 125 GeV with either $H$ or $R$.
To cover both these, we present two scenarios based on the conventions
defined in the previous section.  In the first case, we will fix mass
of the mixed higgs state ($m_H = 125$~GeV) and scan over the mass of
the mixed radion state ($m_R$) from 110 to 600 GeV. Exactly the
opposite is done in the other case.  We describe our analysis using
the first case with the understanding that the identical arguments
apply when $m_R$ is held fixed at 125 GeV.  To improve the efficiency
of our scan, we restrict it to two parameters viz.\ $(m_R,\xi)$ and
take snapshot values of $\Lambda_\varphi$ at 1.5, 3, 5 and 10 TeV.

While it is possible to constrain $\Lambda_\varphi$ further using
either heuristic arguments or from searches for KK excitation of the
RS graviton \cite{Tang:2012pv}, we refrain from doing so to examine
whether the current higgs search data can provide a complementary
method for constraining the parameters of the RS model.  Thus we start
our study with the lowest value radion vev at 1.5 TeV.  Taken together
with the mass limits on the first excitation of the RS graviton, this
might imply values of the bulk cosmological constant well into the
trans-Planckian region where quantum gravity effects may in principle
invalidate the classical RS solution.  However, it may also be
possible to reconcile a low radion vev with rather large gravition
masses in some extended scenarios, such as one including a
Gauss-Bonnet term in the 5-dimensional action \cite{Kim:1999dq,
  Kim:2000pz, Rizzo:2004rq, Choudhury:2013yg, Maitra:2013cta}.

We simulate the kinematics of the signal (higgs production and decay)
using Pythia 8.160 \cite{Sjostrand:2007gs} and reweighting according
to the changed couplings.  In the region where the second resonance
lies between 122-127~GeV, we use Madgraph 5 \cite{Alwall:2011uj} to
calculate the full cross section for $pp \rightarrow X \rightarrow
WW^{(*)}/ZZ^{(*)}/\gamma\gamma$ to include interference from both
states.  The SM rates are taken from \cite{Dittmaier:2011ti,
  Denner:2011mq}.

\subsection{The overall scheme}
\label{sec:scheme}

In this study, we ask two questions: first, what fraction of the
radion-higgs mixing parameter space survives the observed exclusion
limits on signal strengths in various search channels for the SM
higgs; and second, if a radion-higgs scenario can explain the current
data with a better fit than the SM?

Having framed these questions, we compare the theoretical predictions
with observed data in various channels, namely, $\gamma\gamma$,
$ZZ^{(*)} \rightarrow 4\ell$, $WW^{(*)} \rightarrow 2\ell + MET$, $b
\bar b$ and $\tau \bar{\tau}$.  Each channel recieves contribution
from both of the states $H$ and $R$.  Since the production channels
for both $H$ and $R$ are same as the SM higgs (denoted henceforth as
$h_{SM}$), albeit with modified couplings to SM particles, the
production cross section of a given scalar can be written in terms of
the SM higgs production cross section multiplied by a function of the
modified couplings.  We denote this function by $p_{mode}^{R,H}$,
e.g.\ in the gluon-fusion mode,
\begin{equation}
p_{gg}^R(m) = \left. \frac{\sigma(gg \rightarrow R)}{\sigma(gg \rightarrow h_{SM})} \right|_{m_R=m_h=m} =
\frac{B(R \rightarrow gg)}{B(h_{SM} \rightarrow gg)}
\end{equation}
In general, we expect the acceptance of the cuts to depend on (a) the
production mode, and (b) mass of the resonance.  Let us denote the
acceptance of cuts applied for a candidate mass $m$ by the
experimental analysis in a given channel as
$a(m)_{prod-channel}$. Thus the predicted signal strength at a
particular mass $\mu(m) =\sigma/\sigma_{SM}(m_{h_{SM}} = m)$ in any
given decay channel $c$ is given by
\begin{multline}
\mu(m;c) = \displaystyle\sum\limits_{j = gg, VBF, VH} \left \{ p_j^H \frac{a(m;H)_{j}}{a(m;h_{SM})_{j}}  \frac{ \mathrm{B}(H \rightarrow c )}{\mathrm{B}(h_{SM} \rightarrow c) } \right. \\
 \left. +  p_j^R \frac{a(m;R)_{j}}{a(m;h_{SM})_{j}}  \frac{ \mathrm{B}(R \rightarrow c )}{\mathrm{B}(h_{SM} \rightarrow c) }\right\}
\end{multline}

In this analysis, we will be assuming that the state discovered at the
LHC is the higgs-like $H$ ($m_H = m_{h_{SM}} = 125$~GeV) for the first
case and the radion like state $R$ ($m_R = m_{h_{SM}} = 125$~GeV) for
the second.  Therefore, we expect the acceptances to cancel for one of
the terms but not for the other where the second physical state has a
different mass.  For the rest of this section, we derive the formulae
assuming the first case with the understanding that the expressions
for the second case can be obtained merely by switching $m_R$ and
$m_H$.

For channels where the resonance is fully reconstructible viz.
$\gamma \gamma$, $b \bar b$ and $ZZ^{(*)}$, the analyses use
reconstructed mass to identify the resonance and therefore
contribution from the second state are negligible if the resonance is
narrow.  Furthermore, by restricting the number of jets in the final
state, it is possible to restrict contribution to the dominant
production mode.  Since the Lorentz structure of the couplings of $R$
or $H$ is the same as the SM higgs $h_{SM}$, the acceptances also
factor out.  Therefore, for $\mathrm{h}+0~\mathrm{jets}$, in $\gamma
\gamma$ and $ZZ^{(*)}$ channels, $\mu =\sigma/\sigma_{SM}$ takes the
simplified form
\begin{eqnarray}
  \mu(c)  =  p^H_{gg}\frac{ \mathrm{B}(H \rightarrow c)}{\mathrm{B}(h_{SM} \rightarrow c)} = \frac{ \mathrm{B}(H \rightarrow c) \mathrm{B}(H \rightarrow gg)}{\mathrm{B}(h_{SM} \rightarrow c)\mathrm{B}(h \rightarrow gg)}
\label{eqn:simpleratio}
\end{eqnarray}

However, in the $WW^{(*)}$ channel, the final state is not fully
reconstructible and therefore we need to consider contributions from
both the scalar physical states.  Even on restricting to zero- and
one-jet final states (which are largely due to $gg$ fusion), we still
have
\begin{eqnarray}
\label{eqn:WWstrength}
\mu(m;WW) & = & p_{gg}^H \frac{a(m;H)}{a(m;h_{SM})} \frac{
  \mathrm{B}(H \rightarrow WW )}{\mathrm{B}(h_{SM} \rightarrow WW) } +
p_{gg}^R \frac{a(m;R)}{a(m;h_{SM})} \frac{ \mathrm{B}(R \rightarrow
  WW)}{\mathrm{B}(h_{SM} \rightarrow WW)}
\end{eqnarray}

The branching fraction $R \rightarrow WW^{(*)}$ reaches its maximal
value when its mass passes the threshold $m_{R} = 2 m_W$.  At this
point, the largest contribution to the dilepton final state can come
from decay of $R$ rather than $H$.  Therefore, even with fixed mass of
$H$ at 125 GeV, the presence of another state that can contribute to
the signature results in much stronger bounds on the radion-higgs
mixed scenario.  To estimate the effect of this, we have implemented
the kinematical cuts on the leptons, jets and missing energy as
described by the respective ATLAS~\cite{ATLAS-CONF-2013-030} and
CMS~\cite{CMS-PAS-HIG-13-003} analyses.  We verify that our
simulation of these analyses reproduce the expected number of
signal events for a SM higgs within the errors quoted by the
respective analyses.

In the $\mathrm{h}+2~\mathrm{jets}$ channel, the requirement of two
well-separated jets means the dominant contribution comes to VBF
instead of $gg$ fusion.  However, the gluon-fusion contribution is
still a significant fraction and therefore, the correct estimate would
require simulation of the kinematics of $gg \rightarrow
R(H)+2~\mathrm{jets}$ to high accuracy as well as full detector
simulation.  A possible way out is to use the $gg$-fusion subrtacted
numbers as have been reported by ATLAS.  However, to extract this
contribution the ATLAS analysis uses the estimate of gluon fusion
production for SM higgs as a background which requires, by definition,
to assume the SM.  We have therefore neglected the VBF mode in our
study.

Another important effect arises when the mass of both the scalar
eigenstates is close to each other.  In such cases, the interference
effects cannot be neglected.  We have therefore calculated the full
interference effects when $122 < m_R < 127$~GeV. As we shall see in
the next section, this has important effects both on exclusions as
well as on the global best-fit regions.

In addition, there is the possibility that the branching ratio for the
decay $\varphi^{'} \rightarrow h^{'} h^{'}$ can be substantial in
certain regions of the parameter space, resulting in an enhancement
even in fully reconstructible channels.  Such signals are relatively
suppressed for the $WW^{(*)}$ channel because of various vetos on
aditional leptons and jets.  However they contribute to the $ZZ^{(*)}$
and $\gamma\gamma$ channels where the analysis is by and large
inclusive. We have included this kind of processes whenever the
resultant enhancement is more than 5\% of the direct production rate
i.e.\ $\sigma(pp\longrightarrow \varphi^{'}) \times B(\varphi^{'}
\longrightarrow h^{'}h^{'}) \ge 0.05 \sigma(pp\longrightarrow h^{'})$
for the sake of completeness.

We end this subsection by reiterating the parameters used in our scan.
They are $\Lambda_{\varphi}, \xi$ and mass of either of the mixed
radion state $m_R$ (or the mixed higgs state $m_H$), with the other
fixed at 125 GeV. We use four representative values of
$\Lambda_{\varphi}$, namely 1.5 TeV, 3 TeV, 5 TeV and 10 TeV. $\xi$ is
varied over the entire theoretically allowed region according to the
criteria discussed earlier.

\subsection{Allowed regions of the parameter space}

First, we remember that the experiments have provided 95$\%$ upper
limits on the signal strength in each channel, which can be used to
rule out regions of our parameter space incompatible with observed
data.  For the $\gamma\gamma$ and $ZZ^{(*)}$ channel-based exclusions,
we make use of the simplified formula given in
(Eqn.~\ref{eqn:simpleratio}) for the entire range of $m_{R}$.

The case for $WW^{(*)}$ is more complicated in the region where $m_R$
lies in the range 110 - 160~GeV since contribution from both the
eigenstates are of comparable magnitude. Therefore, we add the
contributions from both states (Eqn.~\ref{eqn:WWstrength}).  For
example, for calculating the cross section at say 150~GeV, we consider
the contribution from $m_R=150$~GeV as well as the contribution from
$m_H=125$~GeV to cuts designed for the 150 GeV analysis.  As $m_R$
approaches 160 GeV, the contribution from the 125 GeV state becomes
smaller and smaller till after 160, it is dominated entirely by $m_R$.
After this point, we continue with the simple ratio treatment viz.
\begin{eqnarray}
  \mu(125;WW) = \frac{ \mathrm{B}(R \rightarrow WW) \mathrm{B}(R
    \rightarrow gg)}{\mathrm{B}(h_{SM} \rightarrow WW)\mathrm{B}(h
    \rightarrow gg)}
\label{eqn:simpleratio2}
\end{eqnarray}

A second source of upper limits comes from demanding that the total
signal strength at 125 GeV does not exceed the upper limit at that
mass.  The cuts based on transverse mass e.g.\ the ATLAS cut on
transverse mass demanding $0.75 m_H < m_T < m_H $ cuts off part of the
contribution from $m_R$ state.  
\begin{eqnarray}
\label{eqn:WWstrength2}
\mu(WW)  =   p_{gg}^H \frac{ \mathrm{B}(H \rightarrow WW )}{\mathrm{B}(h_{SM} \rightarrow WW) } +  p_{gg}^R \frac{a(125;R)}{a(125;h_{SM})}  \frac{ \mathrm{B}(R \rightarrow WW)}{\mathrm{B}(h_{SM} \rightarrow WW)}
\end{eqnarray}

In the ATLAS analysis, the kinematical cuts for higgs search up to
mass of 200 GeV are identical excepting the transeverse mass cut. In
the CMS analysis, the cuts vary continuously with mass. We refer the
reader to the relevant papers \cite{Aad:2012uub, ATLAS-CONF-2013-030,
  CMS-PAS-HIG-13-003} for details of the cuts used.

\subsection{Best fit contours}
\label{sec:chisq}

\begin{table}[tp]
\begin{center}
\begin{tabular}{lccc}
\hline
\textbf{Channel} & \textbf{ATLAS} & \textbf{CMS} & \textbf{Tevatron}\\
\hline
$WW^*$ & $1.0 \pm 0.3 $ & $0.68 \pm 0.20$ & \\
$ZZ^*$ & $1.5 \pm 0.4 $ & $0.92 \pm 0.28$ & \\
$\gamma \gamma$ & $1.6 \pm 0.3$ & $0.77 \pm 0.27$ & \\
$\tau \tau$ & $0.8 \pm 0.7$ & $1.10 \pm 0.41$ & \\
$b \bar b$ (Tevatron) & & & $1.97 \pm 0.71$ \\
\hline
\end{tabular}
\end{center}
\caption{Best-fit values of signal strength used for global fits
  \cite{ATLAS-CONF-2013-014, CMS-PAS-HIG-13-005, Aaltonen:2012qt}. \label{tab:bestfit}}
\end{table}

To answer the second question posed at the begining of
Sec.\ \ref{sec:scheme}, we wish to obtain the best fit values for
$\xi$ and the varying scalar mass ($m_R$ or $m_H$) for each value of
$\Lambda_\phi$. We primarily use data in the $\gamma\gamma, ZZ^{(*)}$
and $WW^{(*)}$ channels, which are the most robust.  We also use $\tau
\bar \tau$ data, however, we find that the error bars for these are so
large its role in deciding the favoured region of the parameter space
is somewhat inconsequential.  For the $b\bar{b}$ final state, we use
data in the associated production channels $WH, ZH$
\cite{Aaltonen:2012qt}.  We do not use the data from LHC in this
channel as its error bars are larger even than the $\tau \bar \tau$
channel and therefore do not restrict any of the parameter space.

To find the best fit, our task is to scan the parameter space and find
the values of $m_{\varphi^{'}}$ and $\xi$ for any $\Lambda_\phi$,
which minimise

\begin{equation}
\chi^2 = \sum_i \frac{(\mu_i - \hat{\mu_i})^2}{\bar{\sigma_i}^2}
\end{equation}

\noindent
where $\mu_i = \sigma/ \sigma_{SM}$ is the signal strength at 125 GeV
as calculated in the {\it i}th channel, $\hat{\mu_i}$ denotes the
experimental best fit value for that channel, and $\bar{\sigma_i}$
being the corresponding standard deviation.  Changing $\xi$ and $m_R$
affect the signal strength of $H$ even though $m_H$ is held fixed at
125 GeV.  Again, we use the simple ratio-based formulae for $\gamma
\gamma$, $ZZ^{(*)}$, $b \bar b$ and $\tau \bar \tau$ (using associated
production instead of gluon fusion for $b \bar b$).  For $WW^{(*)}$,
the formula (Eqn.~\ref{eqn:WWstrength2}) is used.  The data points
used for performing global fit are summarised in
Table~\ref{tab:bestfit}.

The 68\% and 95 \% contours are determined using
\begin{equation}
\chi^2 = \chi^2_{min} + \Delta \chi^2   
\end{equation}
where $\Delta \chi^2$ values corresponding to the confidence levels
for seven degrees of freedom (8.15, 14.1) are used.  Since the
best-fit values reported by the experiments are based on combination
of 7 and 8 TeV runs, we combine our signal strengths at 7 and 8 TeV
weigted by the luminosity.

Since the upper limits are based on signal strength mainly due to the
second resonance whereas the best-fit requires the correct signal
strength at 125 GeV, there may be regions with a small chi-squared
that are already ruled out due to constraints on signal from the
second resonance.  We therefore also perform the best fit in the
region left out after the exclusion limits are applied.  However, to
avoid overconstraining the parameter space, we do not include the
exclusions arising from upper limit on the signal strength at 125 GeV
as given by (Eqn.~\ref{eqn:WWstrength2}) while performing the
chi-squared minimisation.

\begin{figure}[tp]
\begin{center}
\includegraphics[scale=0.5]{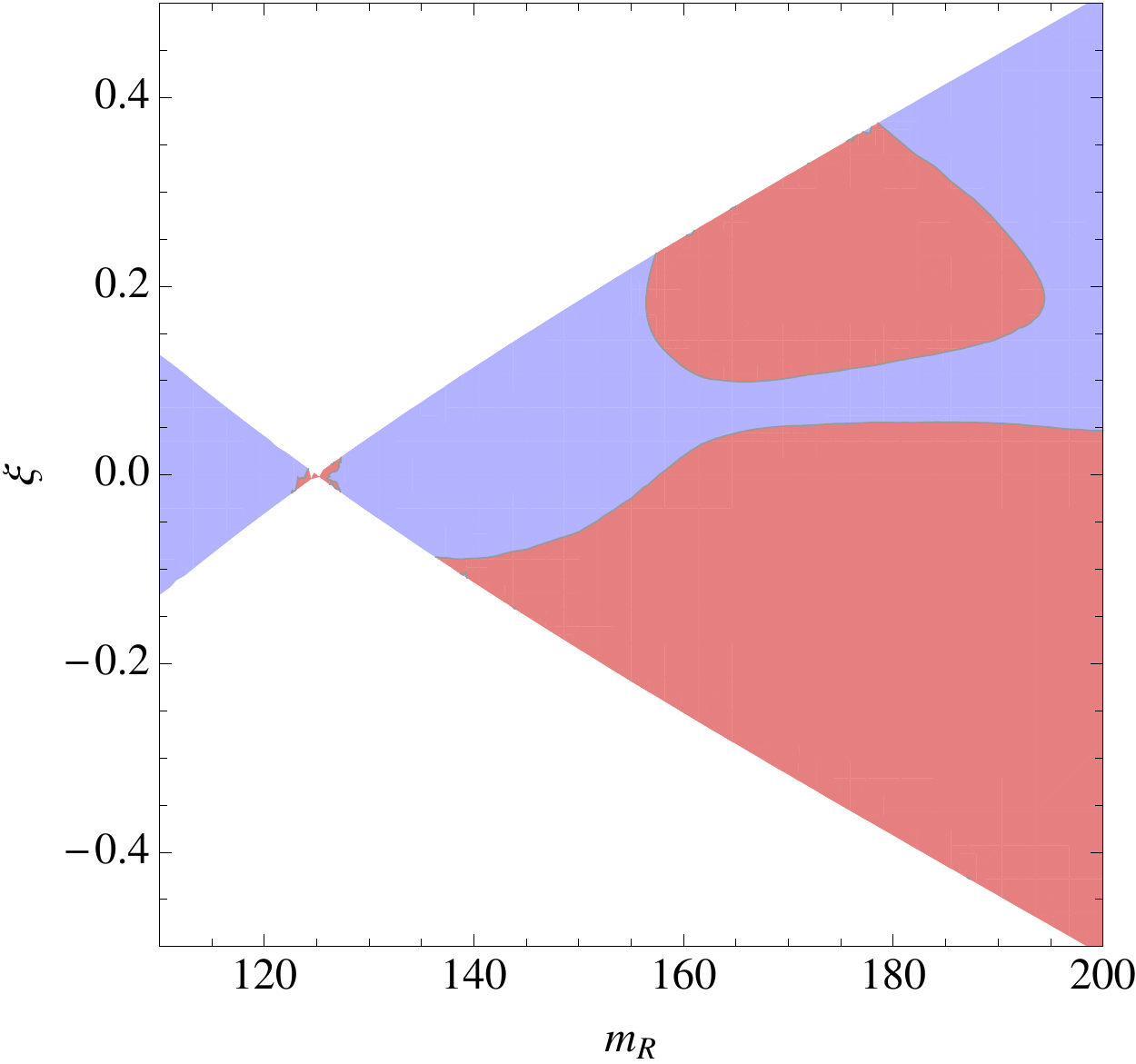}
\includegraphics[scale=0.5]{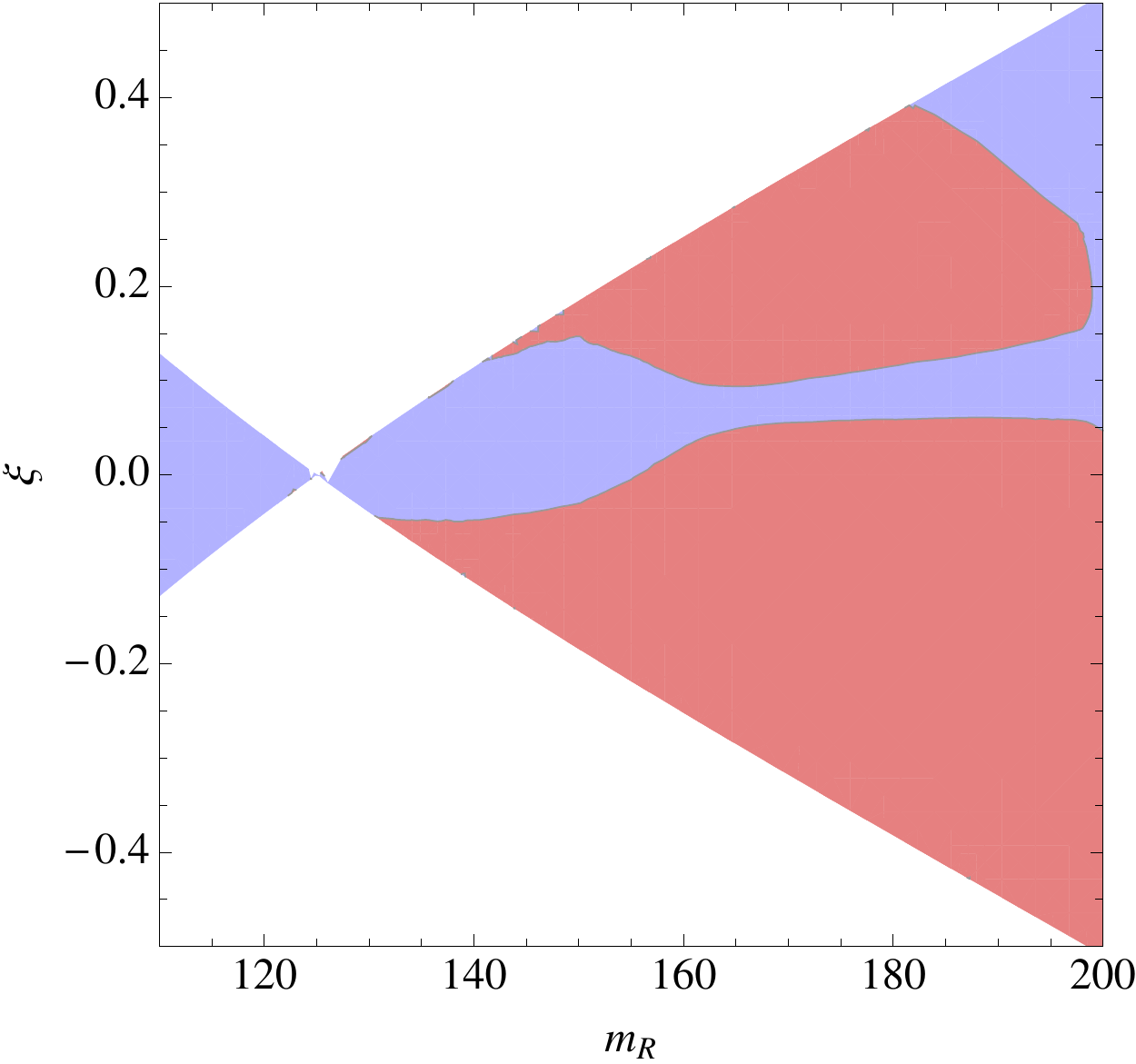}
\includegraphics[scale=0.5]{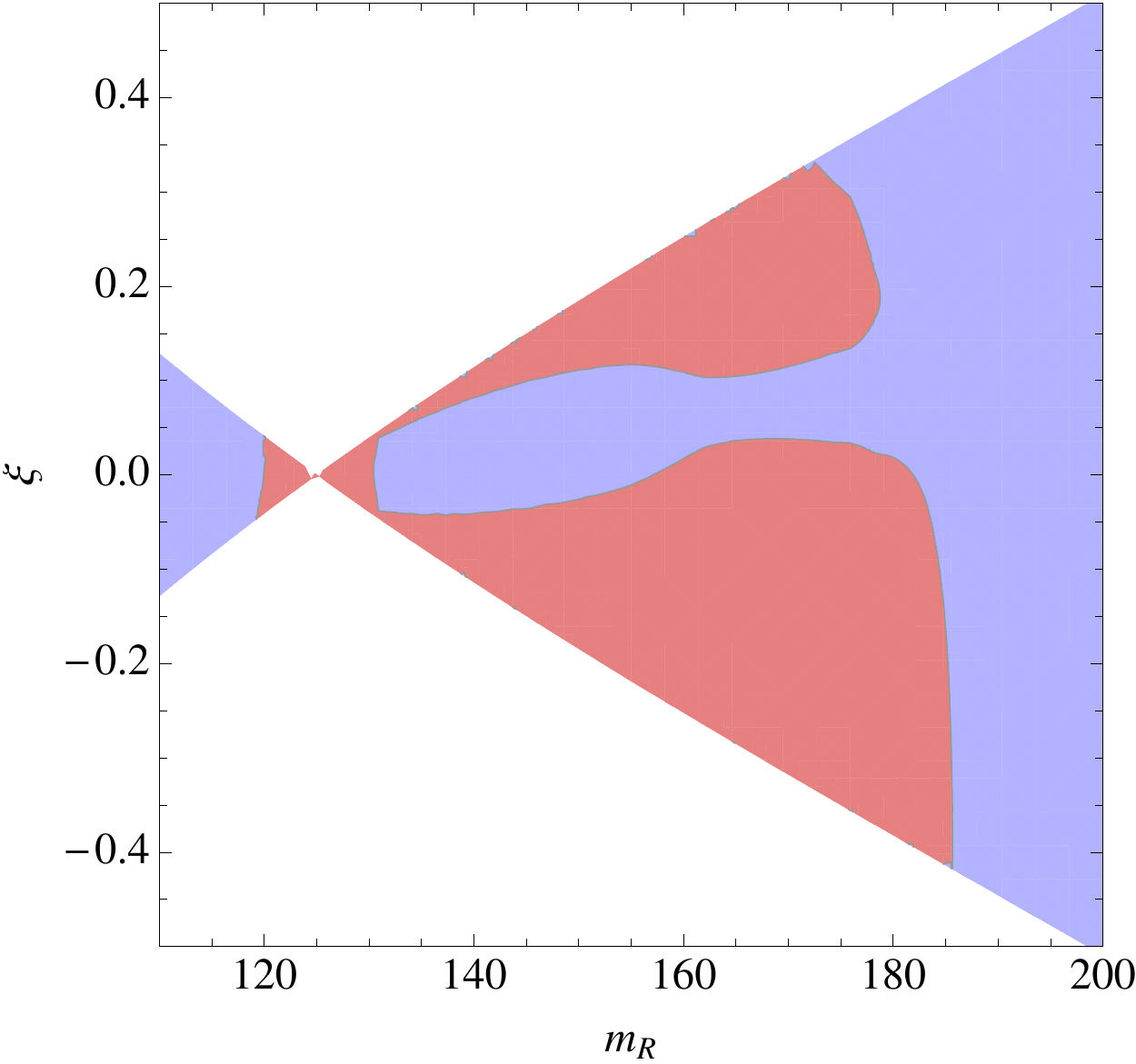}
\includegraphics[scale=0.5]{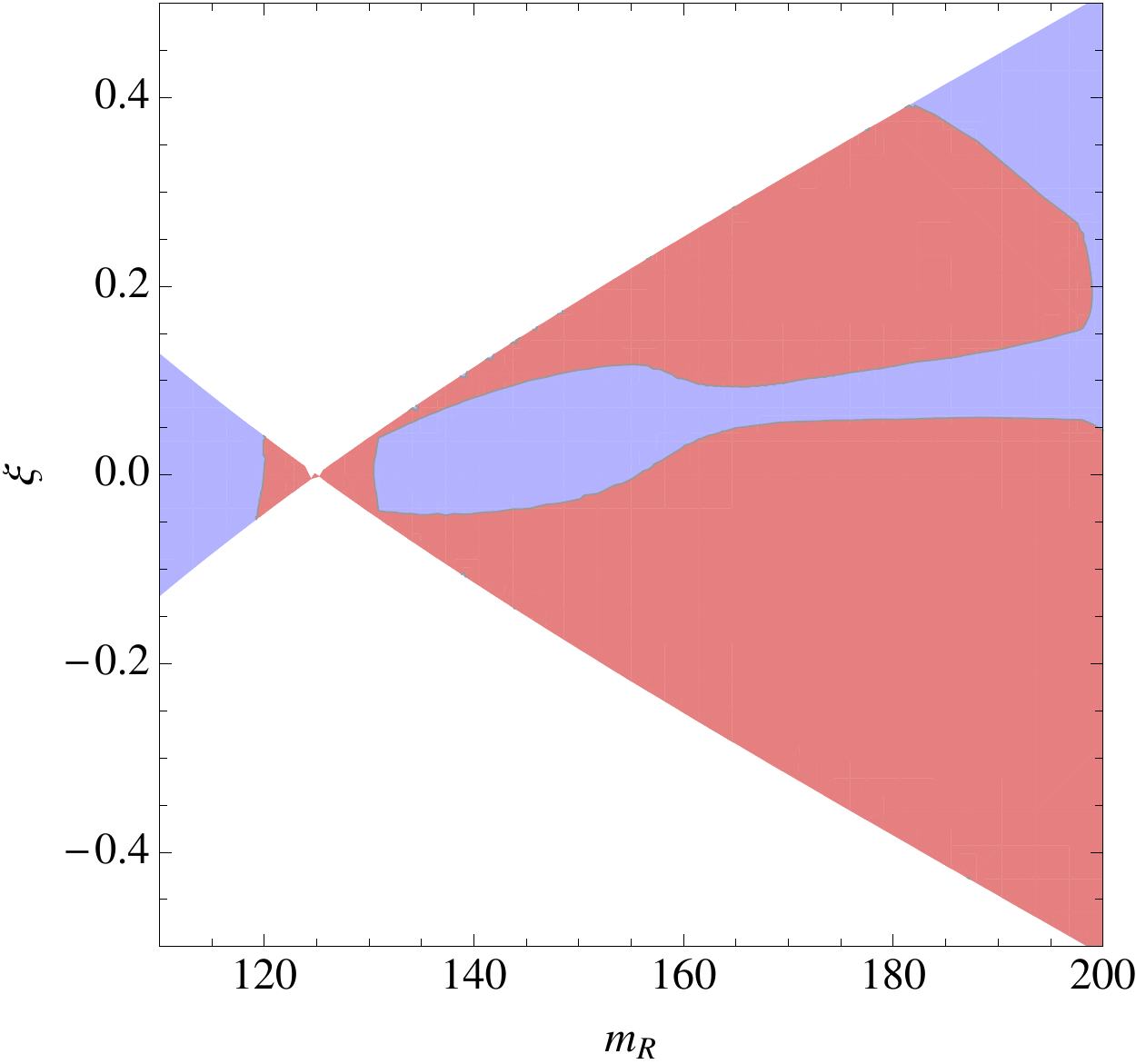}
\caption{\label{fig:WWchange} The effect on the excluded parmeter
  space (shown in red) from various contributions.  The top-left panel
  shows the excluded region using ratios of branching fractions of
  $m_R$ alone. The top-right panel is the exclusion when contribution
  from both states are taken into account.  The bottom-left panel
  shows the exclusion from applying the limit on signal strength at
  125 GeV.  Finally, the bottom-right panel shows the total excluded
  parameter space.  This illustration uses $\Lambda_\varphi = 3$~TeV
  and 95\% CL limits from the ATLAS collaboration.}
\end{center}
\end{figure}

\section{Results and discussions}

The most recent CMS and ATLAS search results exclude the Standard
Model higgs in the mass range $128$ to $600$ GeV at $95\%$
CL~\cite{ATLAS-CONF-2013-014, CMS-PAS-HIG-13-005}.  In this section we
present the regions of the RS parameter space that allow the presence of
an extra scalar consistent with observed upper limits.

We illustrate the effect of taking signal contributions from both
states in Fig.~\ref{fig:WWchange}.  The top-left panel shows the
excluded region when the upper limits are placed on signal strength of
the extra R state alone using only the multiplicative correction of
Eqn.~\ref{eqn:simpleratio}.  This was the approach used
e.g.\ in~\cite{Kubota:2012in}.  However, the presence of two states
means there are two sources of limits --- firstly, we require the
total signal strength at 125 GeV to be less than the observed upper
limit at 125 GeV (bottom-left panel) and secondly, we also require
that the combined signal strength be smaller than the observed limit
at the mass of the radion-like resonance $m_R$ (top-right panel).
Finally we show the effects of both these taken together to give the
full exclusion (bottom-right panel).

\begin{figure}[tp]
\includegraphics[scale=0.6]{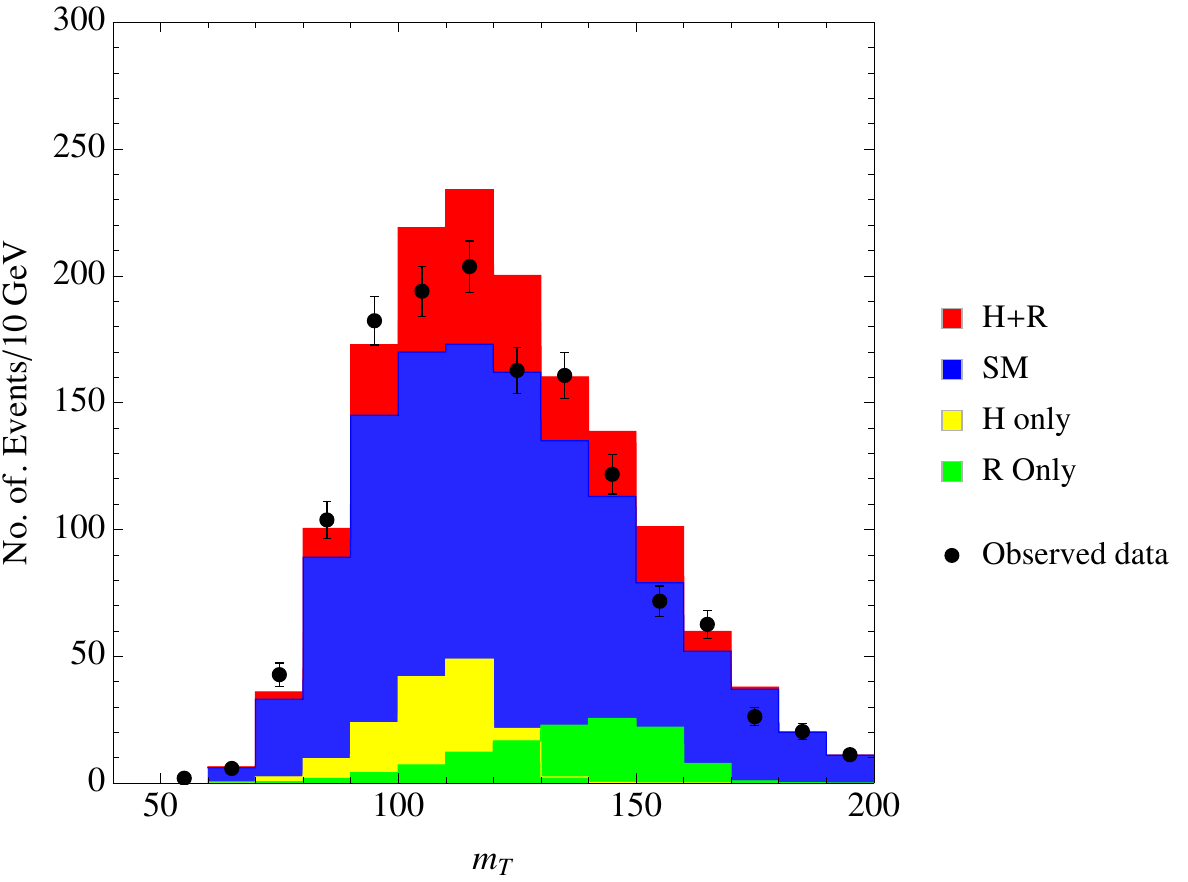} 
\includegraphics[scale=0.6]{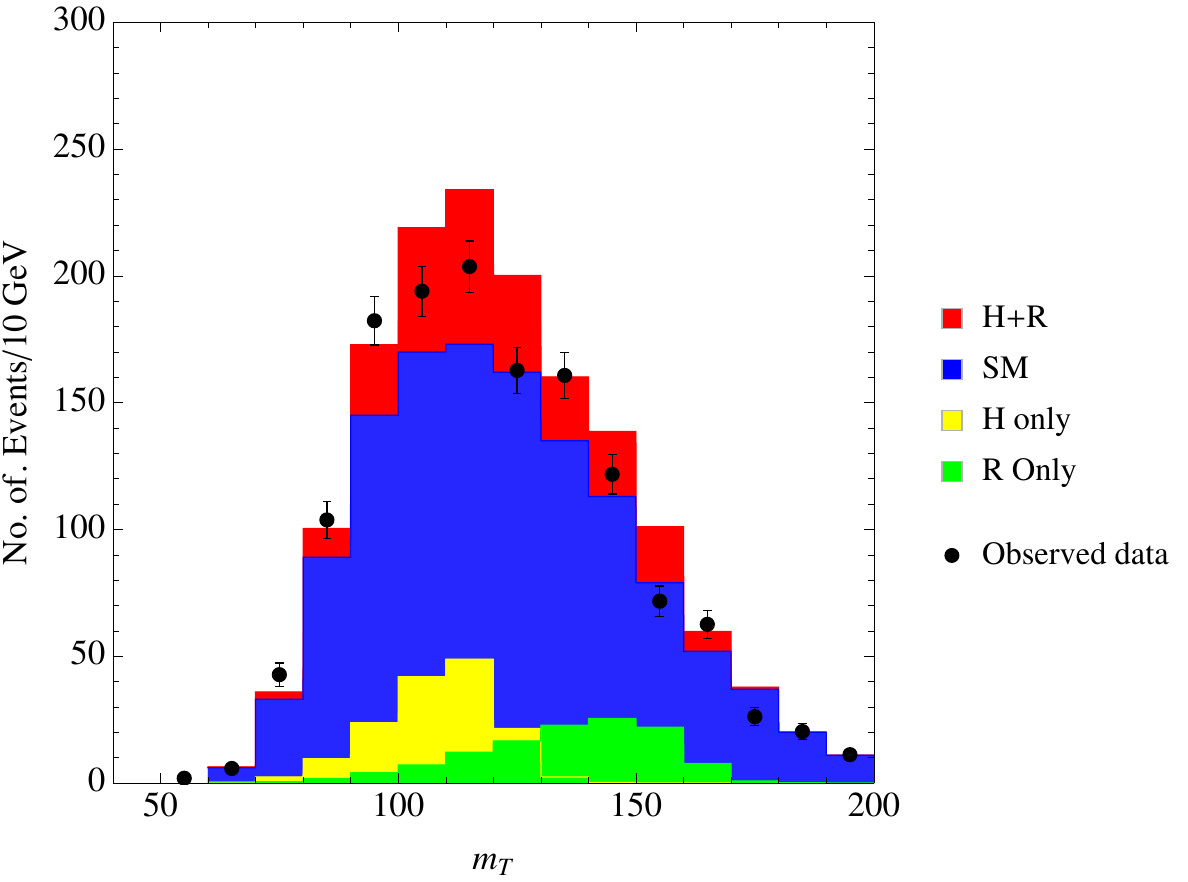} 
\caption{Comparison of $m_{T}$ distribution after contribution from
  both scalars is taken into account for a parameter point that is
  ruled out and one that is not by the ATLAS limits.  The parameters
  for illustration are $\xi = 0.045$ (left; disallowed) and
  $\xi=0.065$ (right; allowed), $m_H=125$~GeV, $m_R=164$~GeV and
  $\Lambda_{\varphi} = 3$~TeV.  The label ``SM'' refers to the total
  SM background as extracted from ~\cite{Aad:2012uub, ATLAS-CONF-2013-030}. \label{fig:mtdist}}
 \end{figure}

A caveat in the above result is that the likelihood function used by
the experiments to place limits makes use of not just on the total
number of events but also the shape of certain distributions like the
lepton invariant mass $m_{\ell \ell}$ or the transverse mass
$m_T$.\footnote{The transverse mass variable is defined as $m_{T} =
  \sqrt{(E_{T}^{\ell \ell} + E_{T}^{miss})^2 - |(p_T^{\ell \ell} +
    E_{T}^{miss})|^{2}}$, where $E_{T}^{\ell\ell}$ is the transverse
  energy of the leptonic system, $p_{T}^{ll}$ is the total transverse
  momentum of the leptonic system and $E_{T}^{miss}$ is the missing
  energy.}

The presence of a shoulder, in e.g.\ the $m_T$ distribution, can be
indicative of a second state and could possibly lead to stronger
exclusions in the region where $m_R > m_H$. For a fixed $\xi$, the
branching fraction $R \rightarrow WW^*$ reaches it's maximum value for
about 160 GeV.  For masses greater than this threshold, the change in
total signal strength is governed mainly by the change the production
cross section.  However, since the production cross section decreases
with increasing $m_R$, the distortion in $m_T$ distribution from the
extra state also becomes smaller with increasing $m_R$ and is maximal
around 160 GeV.

We present the $m_T$ distribution showing extra contribution from $R$
for $m_R = 164$~GeV in Fig.~\ref{fig:mtdist} for two nearby values of
$\xi$ viz. 0.045 and 0.065.  Our calculation of the $m_T$ distribution
is superimposed over the estimated background reported by ATLAS
\cite{ATLAS-CONF-2013-030}.  There are in principle, regions of
parameter space where the contribution at 125 GeV from $R$ even
exceeds that from $H$.  However, we find that the current upper limits
on signal strength in $WW$ channel are so strong that this always
results in a very large total signal strength at $m_R$ and is
consequently ruled out.  This is illustrated in Fig.~\ref{fig:mtdist}
where the point with $\xi=0.045$ shows a significant contribution from
$R$ but we find is already disallowed by the 95\% upper limits on
signal strength at 164 GeV.

This observation justifies our assumption that the distortion in the
$m_T$ distribution is not too large even for $m_R \gsim 160$~GeV.  We
therefore present our results with the assumption that the upper
limits on total signal strength give a reasonably good approximation
of the true exclusion limits even though in principle it corresponds
to a limit on the overall normalisation of the distribution only.

\subsection{Exclusion of the Parameter Space}

\begin{figure}[tp]
\begin{center}
\includegraphics[scale=0.5]{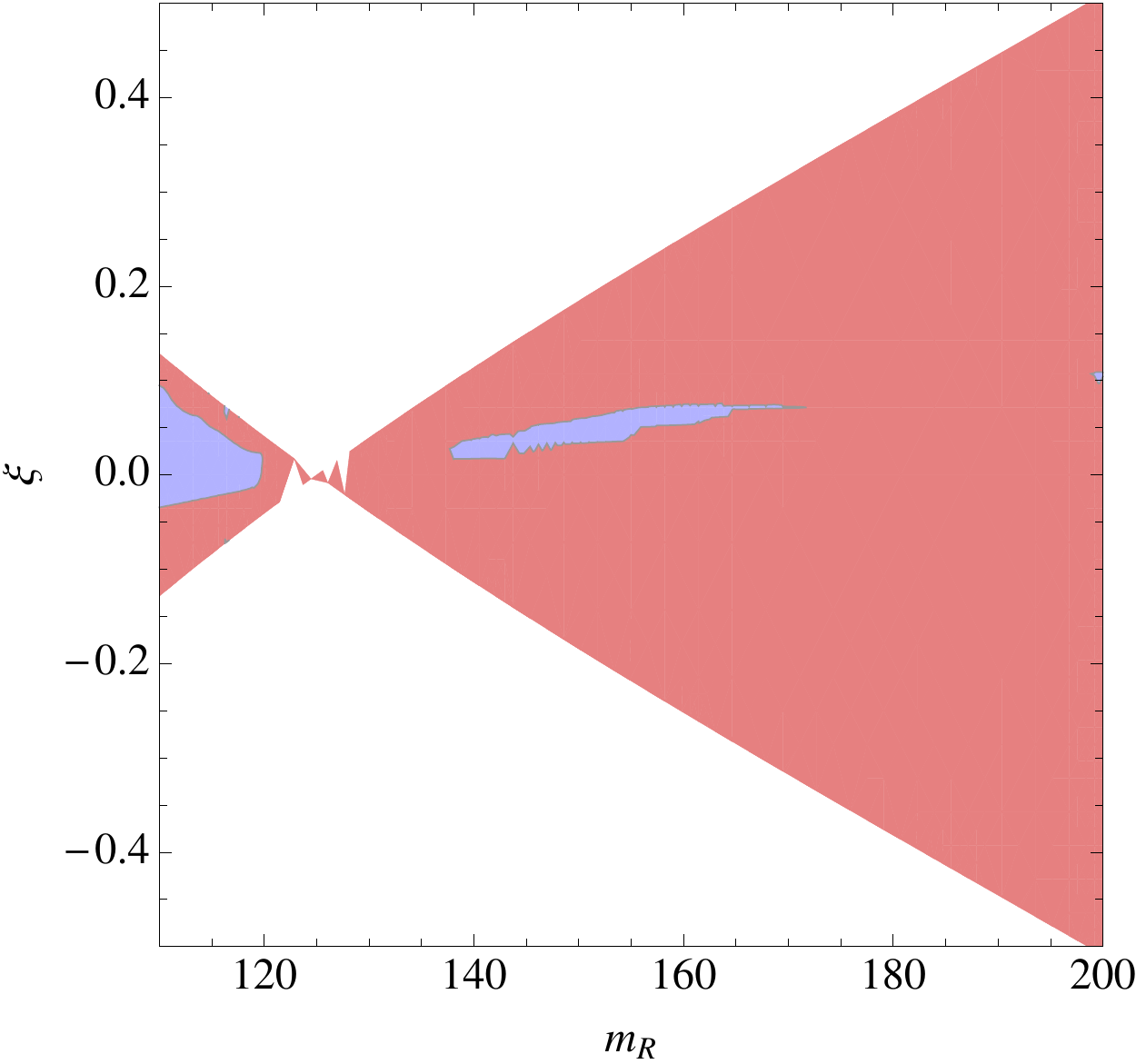}
\includegraphics[scale=0.5]{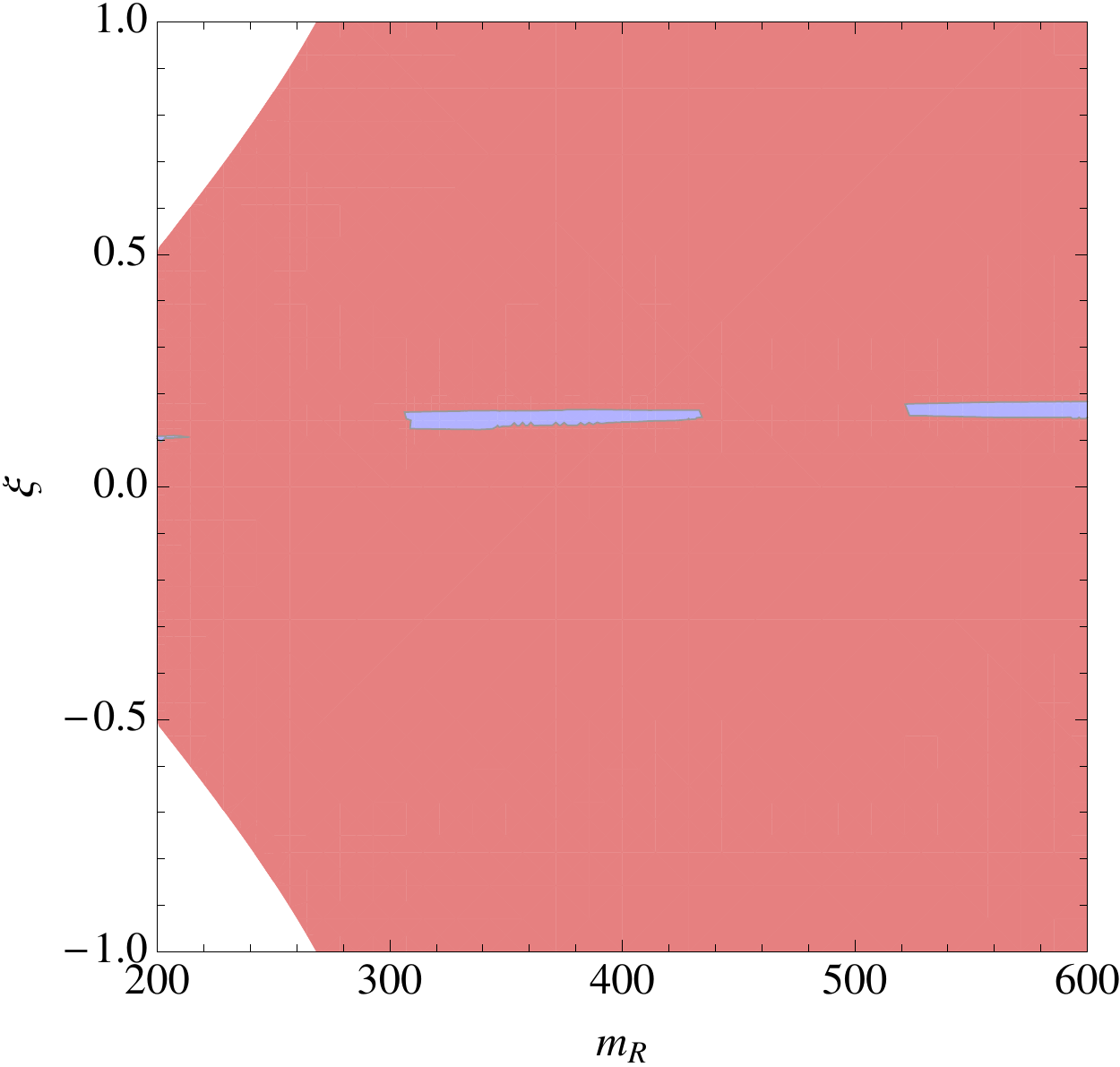}
\includegraphics[scale=0.5]{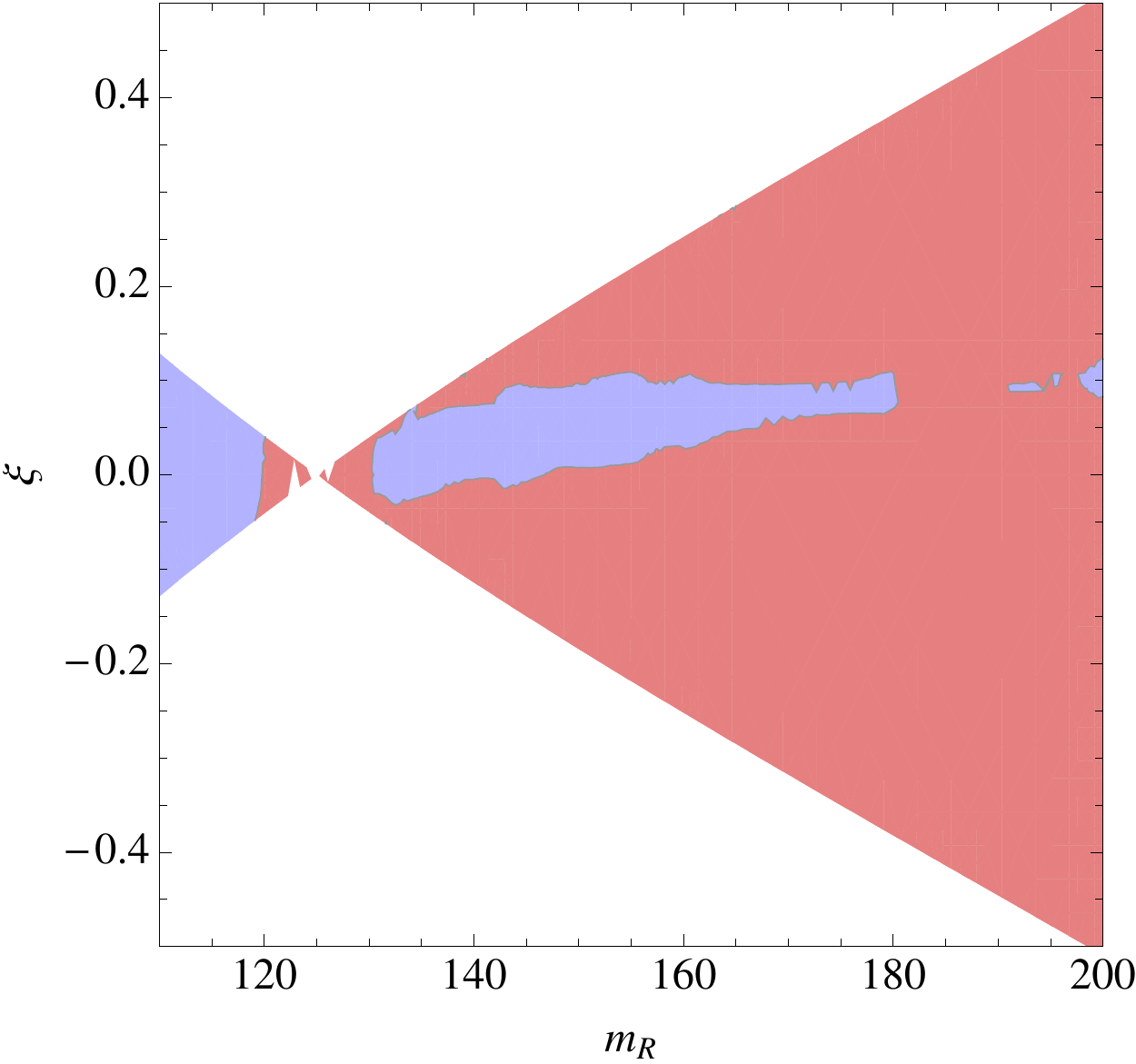}
\includegraphics[scale=0.5]{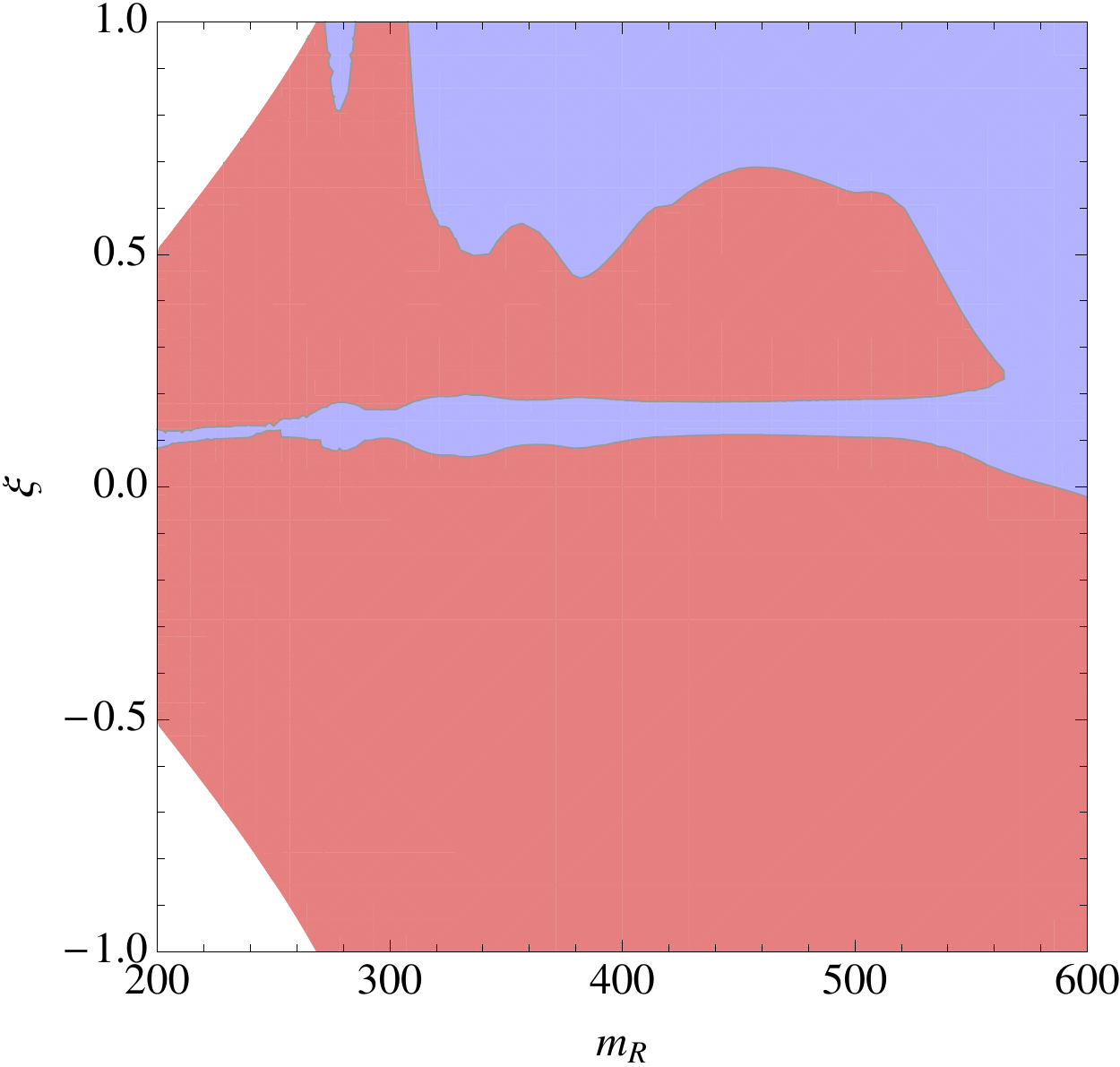}
\includegraphics[scale=0.5]{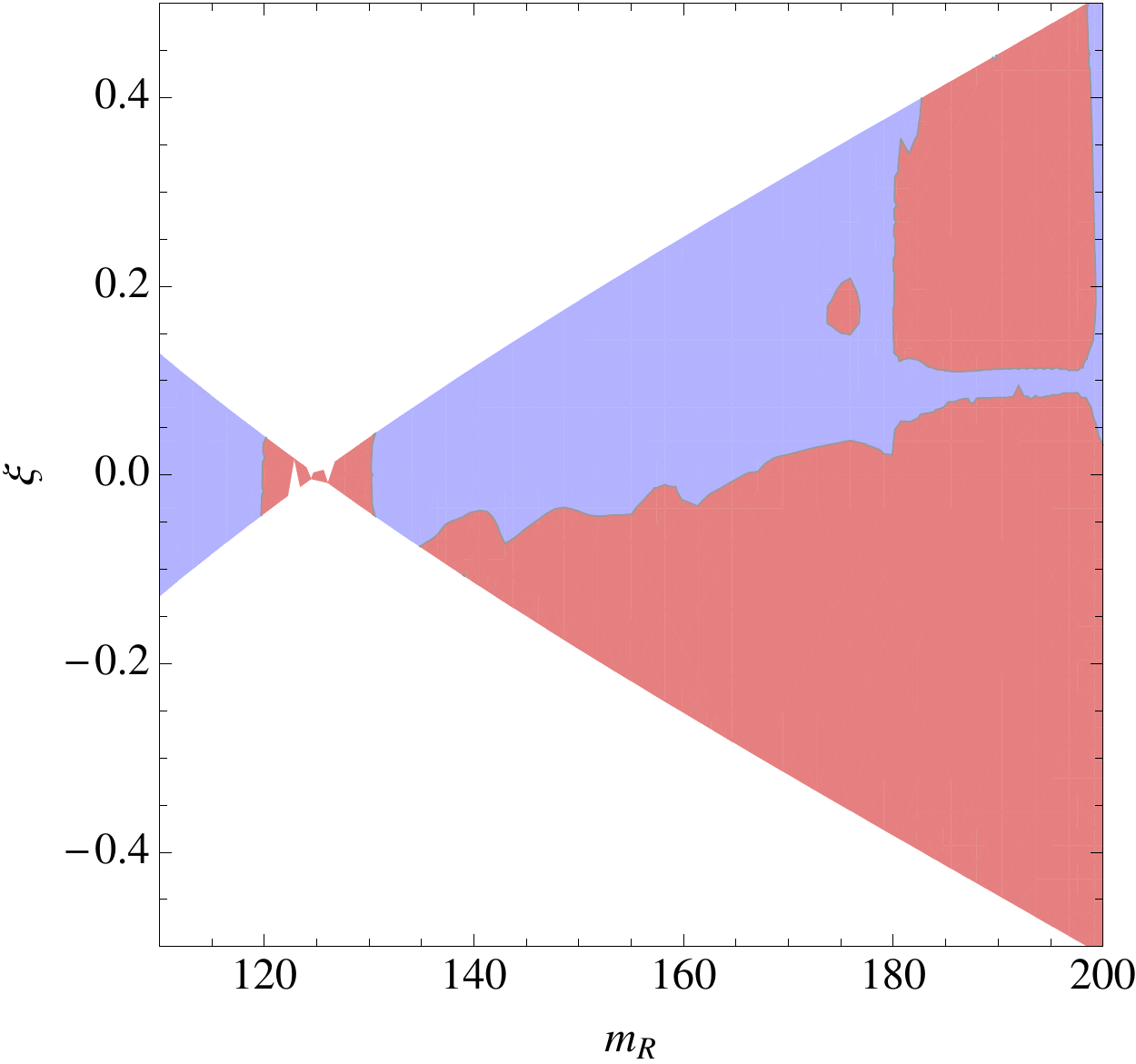}
\includegraphics[scale=0.5]{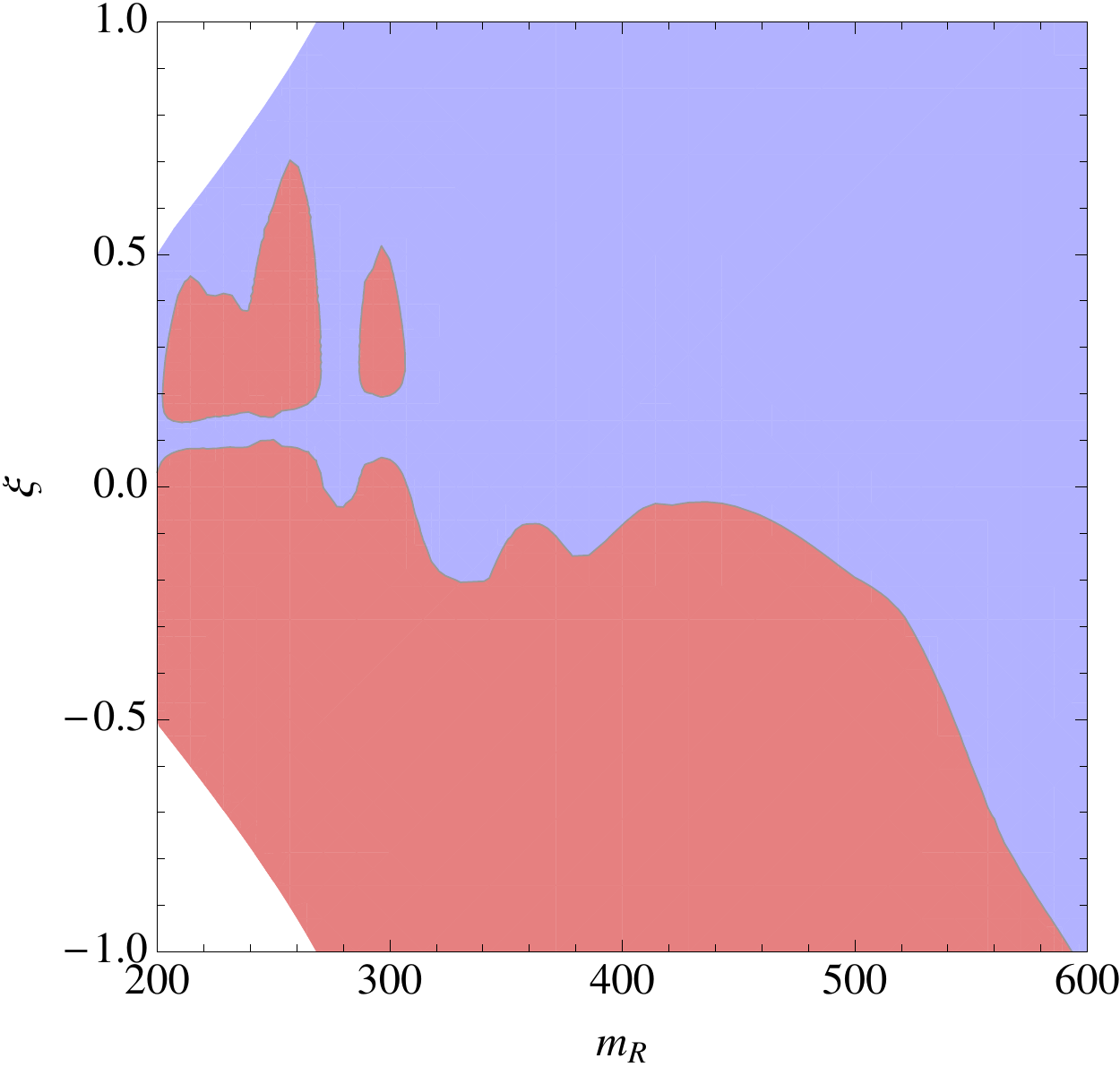}
\caption{\label{fig:exclmH} Excluded parameter space for the case with $m_H = 125$~GeV (shown in red) using 95\% CL limits from the ATLAS and CMS. This illustration uses
  $\Lambda_\varphi = $1.5~TeV(top), 3~TeV(mid) and 5~TeV(bottom).}
\end{center}
\end{figure}

We show the regions of parameter space ruled out from current ATLAS
and CMS data in Fig.~\ref{fig:exclmH}.  As expected, the allowed
parameter space for low $\Lambda_\varphi$ is more restricted than for
higher values.  We find that barring a small sliver close to $\xi =
0$, almost the entire parameter space is ruled out for
$\Lambda_\varphi = 1.5$~TeV.  For $\Lambda_\varphi = 3,~5$~TeV, the
exclusion is less severe.  However, the region with nearly degenerate
$R$ and $H$ states is ruled out.  At large $m_R$, the most stringent
limits come from $ZZ$.  We therefore find regions where a significant
branching fraction $R \rightarrow t \bar t$ reduces the constraints
after $m_R > 350$~GeV. However limits are still restrictive for
negative $\xi$ values as the production via gluon fusion is enhanced
in this region.

We also find that CMS constraints are much stronger than ATLAS.  This
is expected in $WW^{(*)}$ since CMS has provided limits based on the
full 7 and 8 TeV dataset whereas ATLAS has provided only partial
results \cite{CMS-PAS-HIG-13-003, ATLAS-CONF-2013-030}.  We list here
the corresponding conference notes from ATLAS that have been used for
determining the ATLAS limits.  Both experiments give limits in $ZZ$
channel based on the full dataset \cite{CMS-PAS-HIG-13-002,
  ATLAS-CONF-2013-013}.

The $\gamma \gamma$ limits are available only in the range 110-150 GeV
\cite{CMS-PAS-HIG-13-001, ATLAS-CONF-2012-168}, presumably since the
SM higgs decays into the diphoton channel becomes negligibly small
beyond this range.  However, since there can be enhancements to this
rate in the radion-higgs mixed scenario, it may be useful to have the
limits in the full range.  Taking interference of both states when
their masses lie between 122 and 127 GeV pushes the predicted signal
strength beyond the observed upper limits thus ruling out the
degenerate region entirely.  The $b\bar{b}$ limits, from ATLAS, CMS or
Tevatron are found to not affect the extent of the region of
exclusion.

Whenever the limits are based on combined datasets, we combine our
calculated signal strength at 7 and 8 TeV with the luminosities
serving as weights.  For $\Lambda_\varphi = 10$~TeV, we do not find
any significant exclusions.
\begin{figure}[tp]
\begin{center}
\includegraphics[scale=0.5]{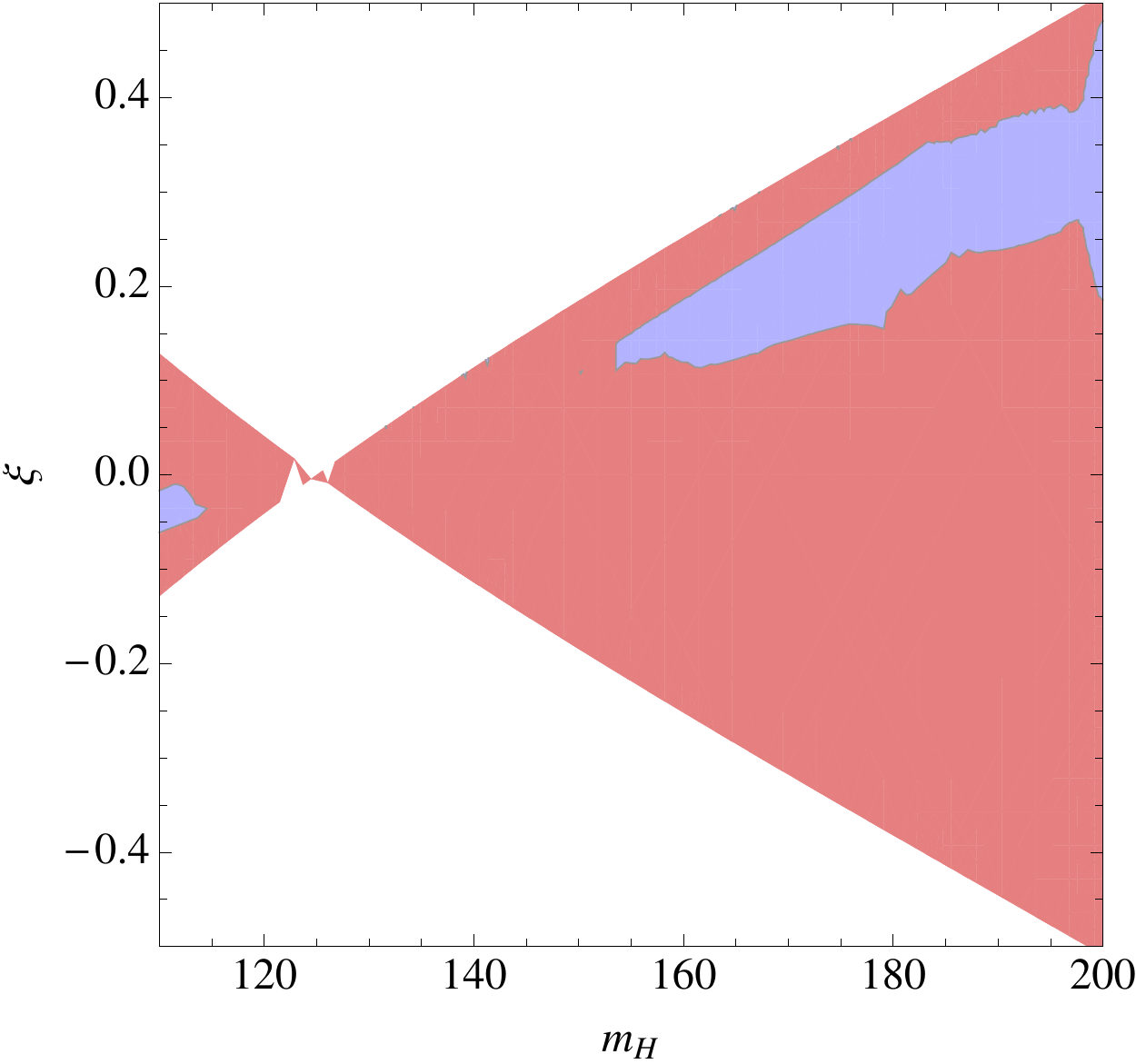}
\includegraphics[scale=0.5]{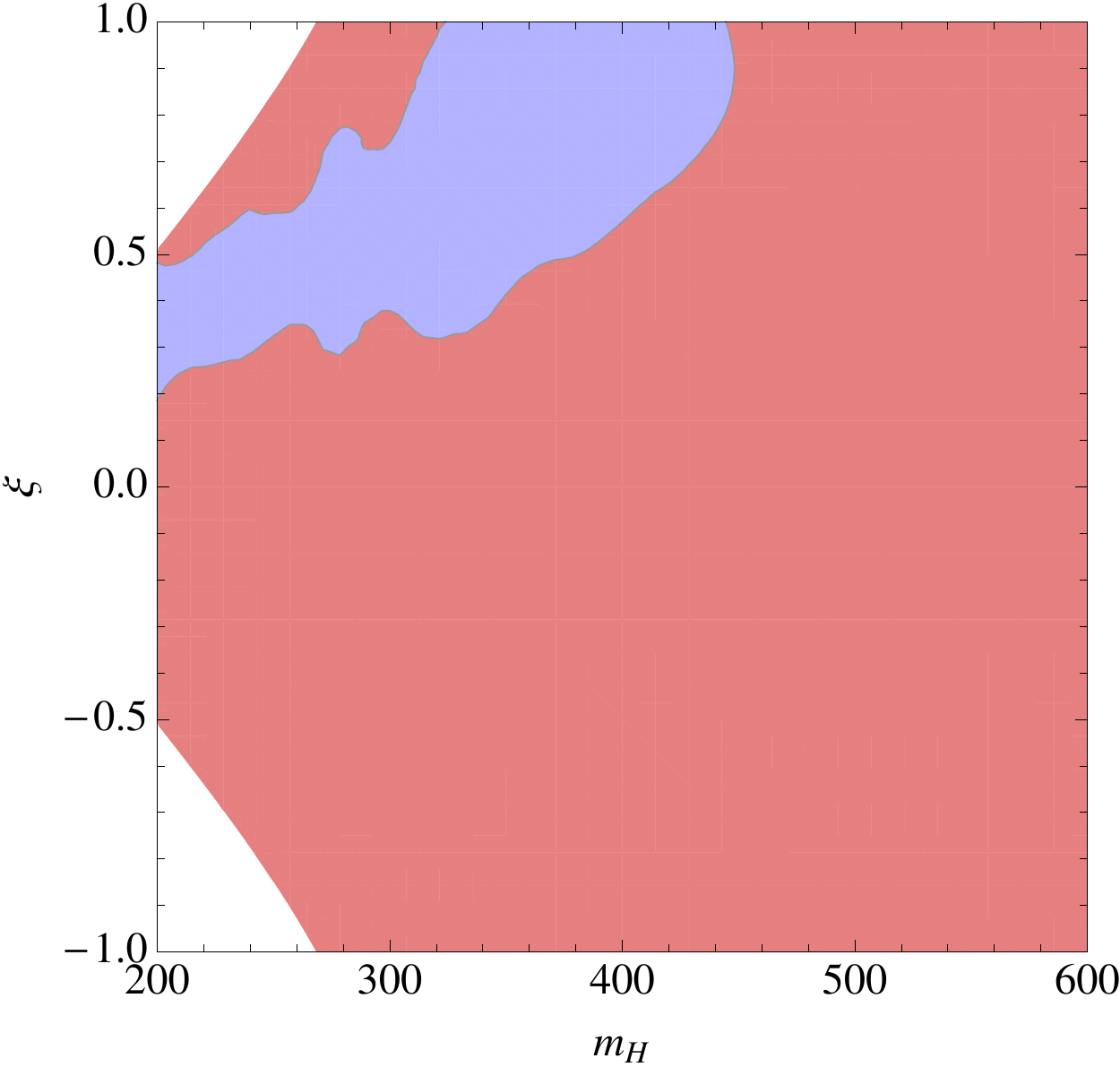}
\includegraphics[scale=0.5]{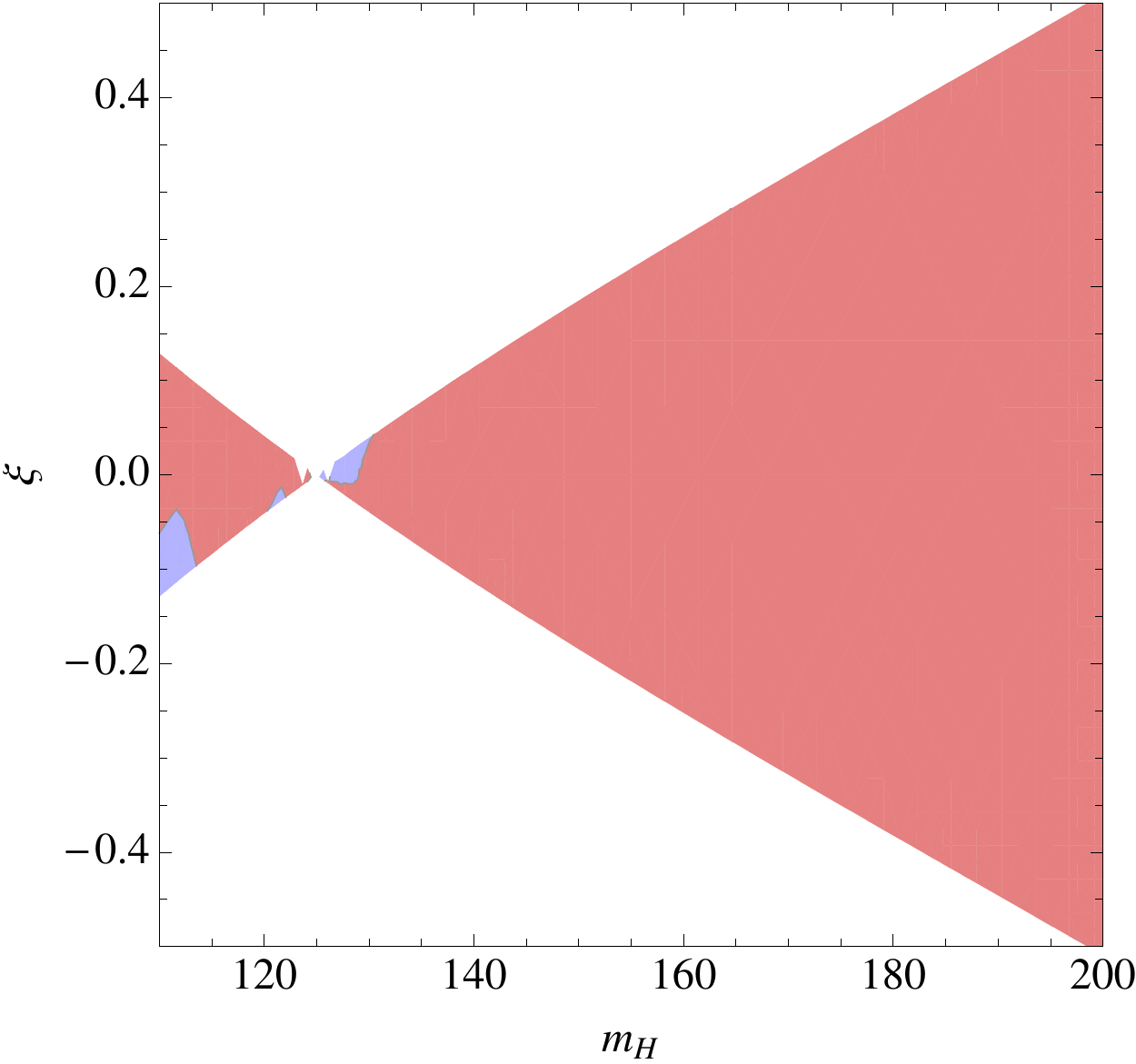}
\includegraphics[scale=0.5]{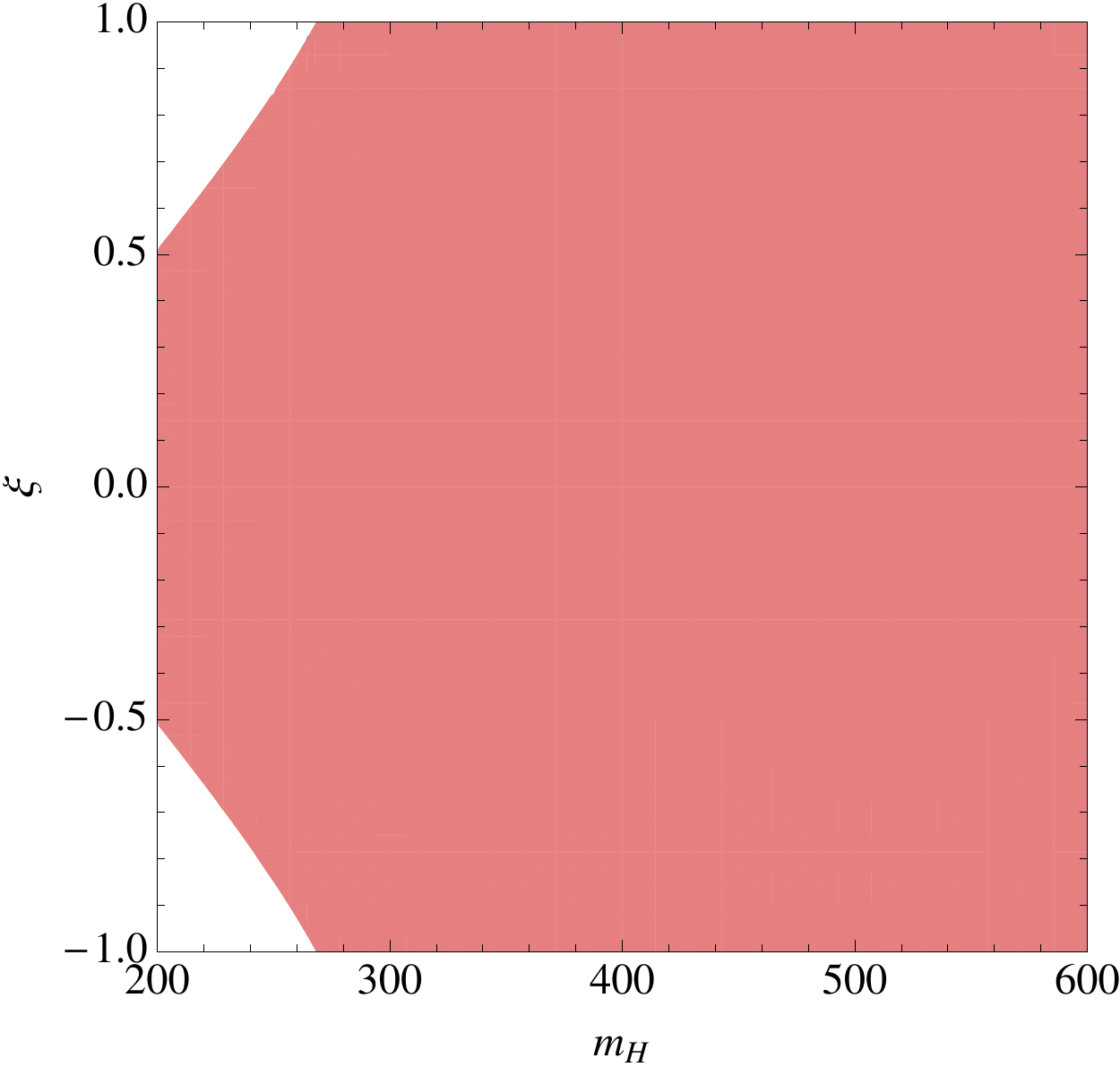}
\caption{\label{fig:exclmR} Excluded parameter space (shown in red)
  for the case with $m_R = 125$~GeV using 95\% CL limits from the
  ATLAS and CMS. This illustration uses $\Lambda_\varphi =
  $1.5~TeV(top) and 3~TeV (bottom). Almost the entire parmeter space
  is excluseded for $\Lambda_\varphi = $5~TeV and higher.}
\end{center}
\end{figure}

A natural question to follow this analysis is what happens if the
boson found at 125 GeV is the $m_R$ state and not the $m_H$ one.  The
exclusions resulting from reversing our analysis in accord with this
change is shown in Fig.~\ref{fig:exclmR}.  We find here that larger
values of $\Lambda_\varphi$ have larger exclusions with almost the
entire parameter space being excluded for $\Lambda_\varphi >
5$~TeV. This is in accordance with \cite{Barger:2011hu} where they
show that a pure radion at 125 GeV is already ruled out.  As
$\Lambda_\varphi$ increases, $H$ becomes more and more like the SM
higgs (and equivalently $R$ becomes a pure radion).  As the lmits on
SM higgs already rule it out in most of the mass range, we find that
nearly the entire parameter space is ruled out too.  In performing the
reverse analysis, we have not considered the interference from both
states, therefore the small allowed region near 125 GeV should be
taken with a pinch of salt.  Since the result should not change from
the earlier case as $m_R \simeq m_H$ in this region and we may assume
that it will be ruled out if a full calculation with interference is
made.

\subsection{Regions of best-fit with the data}
 
\begin{figure}[tp]
\begin{center}
\includegraphics[scale=0.5]{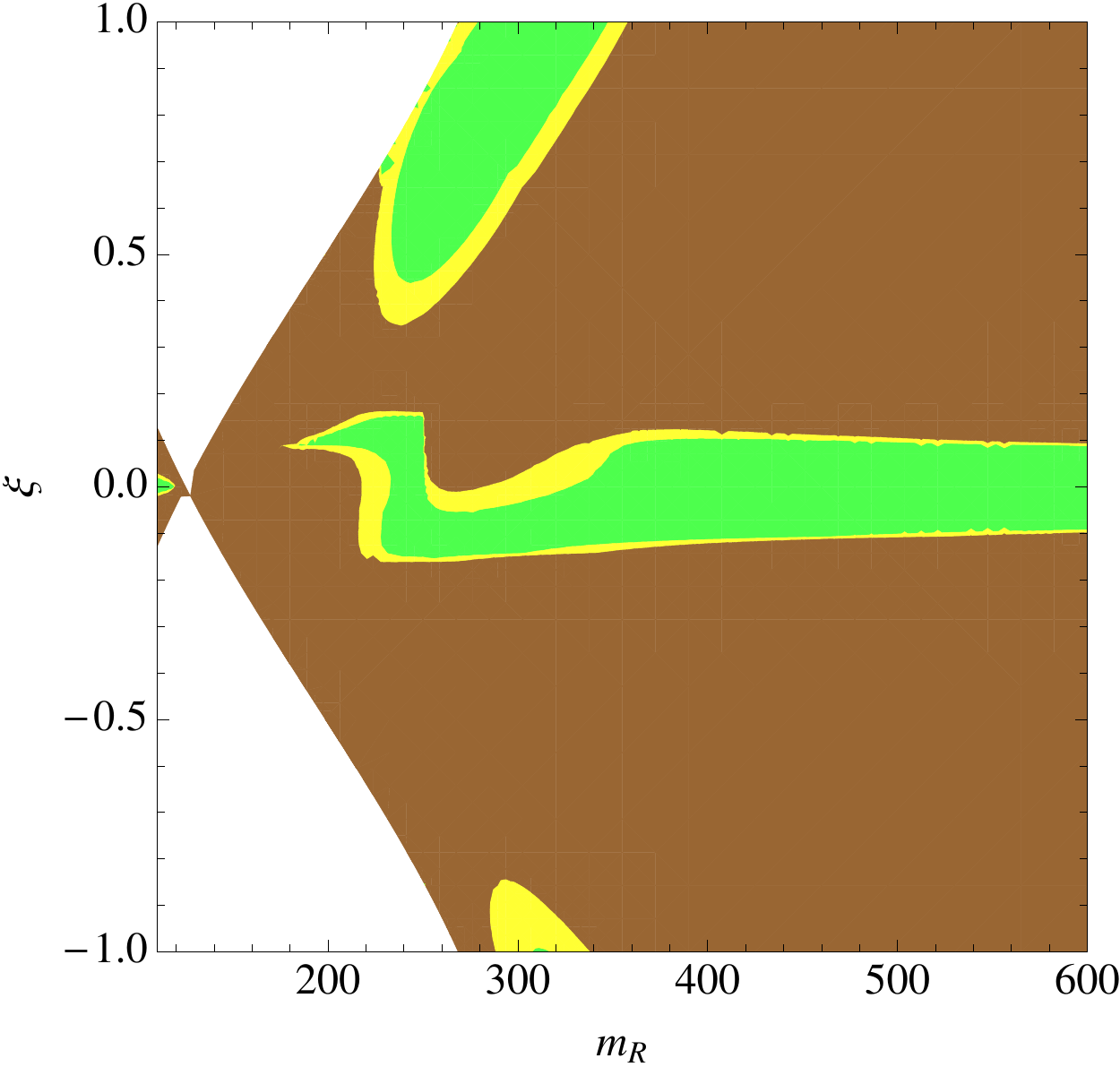}
\includegraphics[scale=0.5]{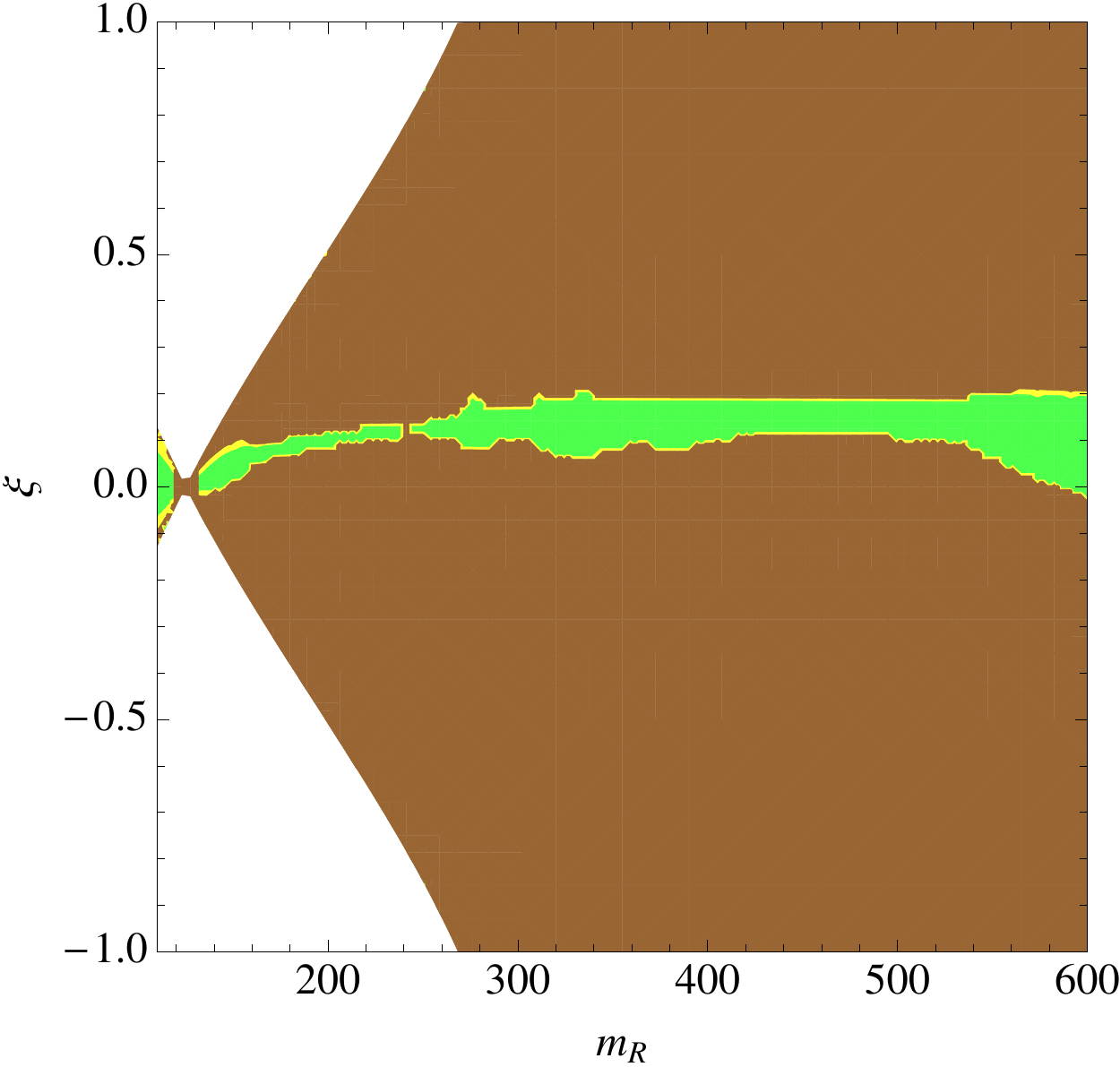}
\includegraphics[scale=0.5]{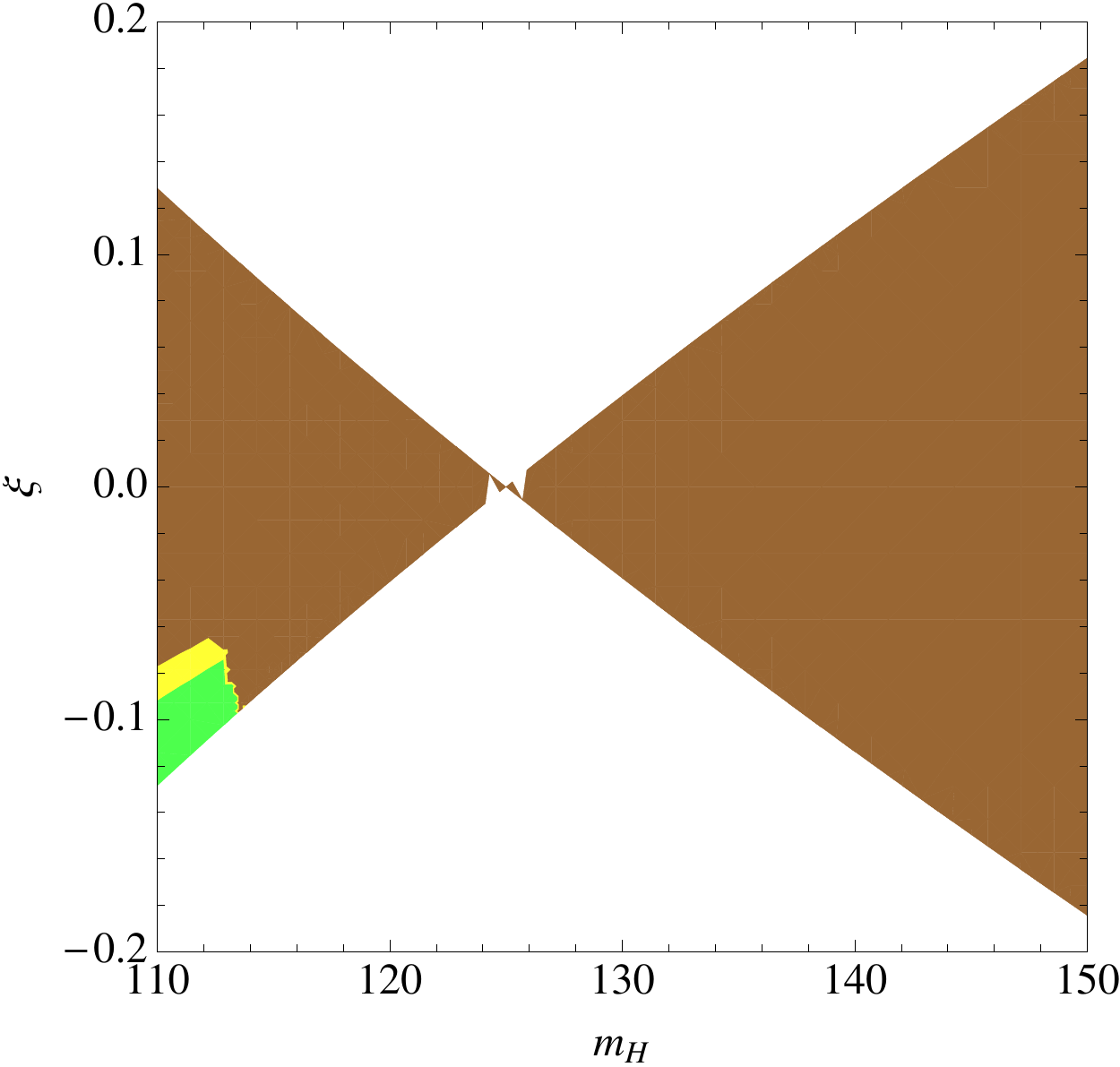}
\caption{\label{fig:BFplots} Regions that agree with current data
  within 68\% (green) and 95.4\% (yellow) for $\Lambda_\varphi =
  3$~TeV. The top-left plot shows the case where no exclusions have
  been taken into account.  The top-right side shows the change after
  taking exclusions into account.  The bottom plot is for the case
  where we hold $m_R=125$~GeV instead of $m_H$.}
\end{center}
\end{figure}

Using the chi-squared analysis outlined in the Sec.~\ref{sec:chisq},
we perform a global fit using the values of signal strength shown in
Table~\ref{tab:bestfit}.  We also perform the same excercise after
removing the regions excluded by the upper limits.  Of course, while
doing so, we do not apply the upper limit on signal strength at 125
GeV.  So the only exclusions considered are those resulting from
limits on signal from $m_R$ only.  For illustraion, we show the
results at $\Lambda_\varphi = 3$~TeV in Fig.~\ref{fig:BFplots}.  The
first panel shows the regions that agree with the data within 68\% and
95\%.  The second panel shows the reduction in the best-fit region
when the exclusions reported in Fig.~\ref{fig:exclmR} are imposed as
well.  The bottom panel shows the best-fit region after exclusions for
the reverse case where $m_R=125$~GeV and $m_H$ is varied.

The chi-squared value for the SM is 10.93 for nine degrees of freedom.
We find that in the first case with $m_H=125$~GeV, there is always a
small region of parameter space that fits with a similar $\chi^2/dof$
as the SM.  For $\Lambda_\varphi = 1.5$~TeV, the minumum chi-squared
value found is 9.06 without exclusions and 11.57 with exclusions at
point $m_R = 600$~GeV and $\xi=0.15$ (after excl.).  For 3 TeV, the
numbers are (9.03, 9.08) respectively with the best-fit point at $m_R
= 407$~GeV and $\xi=0.15$ and for 5 TeV, they are (9.03, 9.04) with
the best-fit point at $m_R = 383$~GeV and $\xi=-0.25$.  Thus, the
exclusions affect less and less as we increase $\Lambda_\varphi$,
which is expected as the excluded parameter space also reduces. In
particular, as the exclusions on negative $\xi$ are relaxed, these
values seem to give a slightly better fit.  Altough, as seen from the
change in $\chi^2$ with and without exclusion, the distribution is
rather flat for large $m_R$.  Also, as the best-fit value for $m_R$ is
at the edge of our scan for $\Lambda_\varphi = 1.5$~TeV, it is
possible that the fit would be further improved by increasing $m_R$.
For larger values of $\Lambda_\varphi$ however, increasing $m_R$ seems
to increase the $\chi^2/dof$ slightly.

The chi-squared for the reverse case is decidedly worse than in the
normal case.  We find that the minimum values of chi-squared after
exclusions are 35.6, 18.22, 52.0 for (1.5, 3, 5 TeV).  Therefore, we
can say that this scenario is strongly disfavoured compared to the SM.

\section{Conclusions}

We have examined the possibility that the currently observed scalar is
one of the two states of a mixed radion-higgs scenario.  To perform
this analysis, we have considered the contribution from both states in
the $WW^{(*)}$ channel, differently affected by cuts, to calculate the
signal strength.  We also take into account effects of intereference
when both states are nearly degenerate.

We find that if the 125 GeV state is radion-dominated, only a very
small region of the parameter space with a small $\Lambda_\varphi$ is
consistent with current upper limits. Even in these regions, the
goodness of fit with data is decidedly worse than in the SM.
Therefore, we may conclude that the idea that the discovered boson at
125 GeV is dominantly radion-like is largely disfavoured.

The second possiblity, namely that the LHC has found a 125 GeV
higgs-dominated scalar, but a radion-dominated state, too, hangs
around to contribute to the observed signals (especially the
$WW^{(*)}$ signal), can not be ruled out with current data.  We find
the scenario with small (but non-zero) mixing and an accompanying
radion-dominated state with high mass results in a good fit for almost
all values of $\Lambda_\varphi$.  However, if we include exclusions on
the presence of the second, radion-dominated boson that would surely
accompany the higgs-dominated state, the goodness of fit is reduced
for TeV-range values of $\Lambda_\varphi$. We find that for
$\Lambda_\varphi$ up to 5 TeV, the SM still provides a better fit.  As
a special case, we find that situations where the two mass eigenstates
are degenerate enough to warrant the inclusion of interference terms,
are ruled out.  Finally $\Lambda_\varphi=10$~TeV is mostly
indistinguishable from the SM as the modifications to signal strengths
are too small to be significant.

\section{Acknowledgements}
We would like to thank Soumitra SenGupta for helpful discussions.  UM
would like to thank Taushif Ahmed, Shankha Banerjee, Atri
Bhattacharya, Nabarun Chakraborty, Ujjal Kumar Dey, Anushree Ghosh,
Sourav Mitra , Manoj Mandal, Tanumoy Mandal and Saurabh Niyogi for
discussions and assistance.  We would also like to thank the
Indian Association for the Cultivation of Science, Kolkata for
hospitality while this study was in progress. ND is supported partly
by the London Centre for Terauniverse Studies (LCTS), using funding
from the European Research Council via the Advanced Investigator Grant
267352. UM and BM are partially supported by funding available from
the Department of Atomic Energy, Government of India for the Regional
centre for Accelerator-based Particle Physics, Harish-Chandra Research
Institute. Computational work for this study was partially carried out
at the cluster computing facility in the Harish Chandra Research
Institute (http:/$\!$/cluster.hri.res.in).

\bibliography{hrad}
\end{document}